\documentclass[11pt]{article}

\usepackage[preprint]{acl}

\usepackage{times}
\usepackage{latexsym}

\usepackage[T1]{fontenc}

\usepackage[utf8]{inputenc}

\usepackage{microtype}

\usepackage{amsmath}
\usepackage{amssymb}
\usepackage{graphicx}
\usepackage{booktabs}
\usepackage{multirow}
\usepackage{xcolor}
\usepackage{subcaption}
\usepackage{tcolorbox}
\tcbuselibrary{breakable}
\usepackage{tabularx}
\usepackage{placeins}
\usepackage{booktabs}

\usepackage[T1]{fontenc}
\tcbuselibrary{listings,skins,breakable}
\usepackage{upquote}
\tcbset{
  promptdisplay/.style={
    enhanced,
    colback=gray!3,
    colframe=black!45,
    colbacktitle=gray!15,
    coltitle=black,
    fonttitle=\bfseries\small,
    boxrule=0.4pt,
    arc=1mm,
    left=1mm,
    right=1mm,
    top=1mm,
    bottom=1mm,
    listing only,
    listing options={
      basicstyle=\ttfamily\footnotesize,
      columns=fullflexible,
      keepspaces=true,
      breaklines=true,
      breakatwhitespace=false,
      showstringspaces=false,
      upquote=true,
      aboveskip=0pt,
      belowskip=0pt
    }
  }
}

\title{Is Position Bias in Dense Retrievers Built In---or Learned from Data?}
\author{Anonymous ACL submission}
\author{
  Daegon Yu\thanks{Equal contribution.} \\
  Sionic AI \\
  \texttt{dgyu@sionic.ai}
  \And
  SeungYoon Han\footnotemark[1] \\
  Sionic AI \\
  \texttt{seungyoon@sionic.ai}
  \And
  Woomyoung Park \\
  Sionic AI \\
  \texttt{max@sionic.ai}
}

\begin{document}
\maketitle
\begin{abstract}

Dense retrievers exhibit positional bias, favoring documents whose query-relevant information appears near the beginning and degrading retrieval performance when the information appears later. While prior work on positional bias in dense retrievers has largely focused on architectural explanations, we study how the positional distribution of evidence in training data affects retrieval-level bias direction. To test this, we construct synthetic position-targeted training sets in which query-relevant evidence appears at the beginning, middle, or end of documents, and fine-tune eight architecturally diverse pretrained models under position-skewed and balanced training distributions. At the ranking level, we observe a strong directional pattern across the examined models: skewed training distributions favor evidence at the corresponding positions. Position-balanced training reduces positional sensitivity by 56--87\% on position-aware benchmarks, with competitive mean retrieval performance in our controlled setting. Representation-level analyses further suggest that fine-tuning often reshapes learned positional preferences, although pre-existing architectural or pretraining-specific tendencies persist in some models. These results identify training-position distribution as a major controllable factor in retrieval-level position bias and suggest balanced data curation as a practical mitigation strategy. Our code is available at \url{https://github.com/sionic-ai/position-bias-dense-retrievers}.

\end{abstract}
\section{Introduction}

Dense retrievers \citep{dpr, contriever} now serve as a core component in open-domain question answering and retrieval-augmented generation \citep{rag, adaptive-rag}. Yet they exhibit a systematic \textit{position bias}. Retrieval performance drops substantially when query-relevant information appears in the middle or end of a document rather than near the beginning \citep{dwell, empirical-study}. A retriever that disproportionately favors early positions risks missing critical information, potentially degrading downstream RAG performance \citep{collapse}; understanding the source of this bias is therefore important.

Prior work has largely examined position bias empirically: it has been observed across training stages \citep{dwell}, positional encodings \citep{quantifying}, and pooling-token attention patterns \citep{fairness}. \citet{posir} further show that positional sensitivity does not correlate with architectural factors. The underlying cause in dense retrievers thus remains unclear. In autoregressive transformers, causal attention has been identified as a primary cause of position bias \citep{pine, emergence-of-position-bias}. Yet encoder-based dense retrievers—which lack causal masking—still exhibit strong primacy bias \citep{dwell, empirical-study}, indicating that architectural factors alone may not fully explain position bias in dense retrievers.

This raises a fundamental question: to what extent can retrieval-level position bias be changed by the positional distribution of fine-tuning data, beyond tendencies induced by architecture and pretraining? In this work, we hypothesize that training-position distribution is an important factor in shaping retrieval-level position bias in dense retrievers. Two forms of positional skew motivate this hypothesis: in training corpora, texts such as news articles place key information in early positions \citep{inverted-pyramid, enhanced-news-retrieval}, and in retrieval fine-tuning data, such as MS~MARCO, query-relevant passages are heavily concentrated in early document positions \citep{mitigating-re-ranking, dwell}. Yet no prior work has directly manipulated training data to isolate its role.

To test this hypothesis, we construct position-controlled datasets in which query-relevant information appears at the beginning, middle, or end of documents, and fine-tune eight architecturally diverse pretrained models—covering encoder and decoder architectures, multiple positional encodings, and different pooling strategies—on them. If models with fundamentally different positional processing nevertheless develop bias patterns that mirror the training distribution, this would suggest that architecture alone cannot fully explain the bias. We evaluate positional sensitivity on position-aware benchmarks and standard retrieval performance on evidence-annotated BEIR subsets.

Our key finding is that retrieval-level position bias direction follows the training data distribution across all eight models, despite their architectural differences: begin-skewed data produces begin-favoring retrieval, mid-skewed data produces mid-favoring retrieval, and end-skewed data produces end-favoring retrieval. Position-balanced training reduces positional sensitivity on position-aware benchmarks while preserving competitive retrieval performance, suggesting that data curation can reduce position bias.

Our contributions are as follows:
\vspace{-0.5em}
\begin{itemize}
\item We design a position-controlled data construction pipeline and release the datasets, enabling controlled experiments on the effect of training data on retrieval-level position bias.
\vspace{-0.5em}
\item We show that training data distributions shape the direction of retrieval-level position bias, with controlled experiments on eight architecturally diverse models revealing predictable shifts in bias direction.
\vspace{-0.5em}
\item We show that position-balanced training reduces positional sensitivity while preserving competitive retrieval performance, suggesting that position bias can be reduced through data curation.
\end{itemize}
\section{Related Work}

\paragraph{Position Bias in Dense Retrievers.} Dense retrievers exhibit position bias, favoring evidence at the beginning of documents \citep{collapse, quantifying, empirical-study}. Across retriever types, dense embedding and ColBERT-style models show performance degradation due to this bias, while BM25 and cross-encoder rerankers remain robust \citep{empirical-study}. \citet{posir} evaluate embedding models on a position-aware benchmark and find that most exhibit primacy bias, though positional sensitivity does not correlate with architectural factors—model size, vector dimension, attention mechanism, or pooling strategy. Similarly, \citet{quantifying} report that the bias persists across positional encodings—APE, ALiBi, and RoPE. These findings show that position bias is widespread in dense retrievers, but they do not explain its cause.

\paragraph{Architectural Explanations.} Prior studies have examined architecture-based explanations for position bias in dense retrievers, but they do not fully explain the observed bias patterns. \citet{fairness} link primacy bias to front-loaded self-attention in pooling-token embeddings of encoder-based models, though its generality across the diverse architectures used in dense retrieval has not been established. In autoregressive transformers, by contrast, \citet{emergence-of-position-bias} prove that causal attention favors earlier tokens with deeper layers amplifying the effect, and \citet{pine} show that RoPE favors nearby tokens through distance-dependent attention decay. However, encoder-based dense retrievers, which lack causal masking, still exhibit strong primacy bias \citep{dwell, empirical-study}, and RoPE-based  decoder retrievers such as Qwen3-Embedding show primacy rather than recency bias \citep{empirical-study, posir}, indicating that architectural factors alone do not fully explain position bias in dense retrievers.

\paragraph{Training Data as a Source of Bias.} Training data has also been implicated as a source of position bias. \citet{dwell} show that position bias emerges during unsupervised contrastive pre-training and is amplified by MS MARCO fine-tuning, where relevant passages are disproportionately concentrated in early document positions. Similarly, \citet{collapse} find that MS~MARCO-trained models exhibit stronger position bias than unsupervised Contriever. Earlier work connects training data to position bias in rerankers: \citet{mitigating-re-ranking} show that rerankers trained on data with early-skewed answer positions inherit this bias. Across these studies, training data appears as a common factor, yet the evidence comes from observation rather than direct manipulation of the positional distribution. Our work addresses this gap by training eight architecturally diverse models on position-controlled datasets, providing direct evidence that training data distribution drives the direction of position bias in dense retrievers.

% \subsection{Data-Centric Perspectives in Information Retrieval}

% Our experimental approach---using controlled data manipulation to diagnose model behavior---builds on a well-established tradition in IR. Training data composition has been repeatedly shown to be a primary determinant of dense retrieval performance, from the design of negative sampling strategies \citep{dpr, ance, star-adore, rocketqa} to synthetic data generation \citep{inpars, gpl}. \citet{promptagator} demonstrated this most strikingly: dual encoders trained on just eight annotated examples with LLM-generated queries outperform models trained on over 500K MS~MARCO examples, illustrating that data distribution matters more than data volume.

% Crucially, biases present in training datasets propagate directly to model behavior. \citet{training-induced-bias} showed that source bias in dense retrievers is training-induced rather than inherent, emerging only after MS~MARCO fine-tuning. This establishes a clear precedent: systematic properties of training data produce systematic biases in retrieval models. Yet no prior work has applied this data-centric lens to position bias specifically. Our work does so, treating controlled data manipulation not as a tool for improving effectiveness, but as an analytical instrument for establishing the causal link between training data distributions and position bias.

\section{Method}
\label{sec:method}

Our approach has two components: a data construction pipeline that produces position-controlled training datasets (\S\ref{sec:data-construction}), and an experimental design that tests how changing the positional distribution of fine-tuning data affects retrieval-level position bias (\S\ref{sec:experiment-design}).

\subsection{Position-Controlled Data Construction}
\label{sec:data-construction}

We construct datasets where the location of query-relevant information is controlled through a three-stage pipeline: corpus preparation with length-stratified binning, position-targeted query generation, and multi-reranker position verification.

\subsubsection{Corpus Preparation}
\label{sec:corpus}

We use English Wikipedia as our source corpus for its topical diversity and wide range of article lengths. Within each pool, we stratify articles by character count into five length bins (256–512, 512–1024, 1024–2048, 2048–4096, and 4096–8192), using character count rather than token count for tokenizer-agnostic consistency across models. Each document is divided into three equal-length segments---\textit{beginning}, \textit{middle}, and \textit{end}---following \citet{empirical-study}.

\subsubsection{Position-Targeted Query Generation}

For each document, we generate queries targeting each of the three positional segments using persona-conditioned prompting with GPT-4o-mini\footnote{{\url{https://developers.openai.com/api/docs/models/gpt-4o-mini}}}, following \citet{qwen3-embedding}. A persona is sampled from PersonaHub \citep{personahub} to encourage diverse information needs; the model then generates a query answerable from only the target segment. This yields three query subsets---$q_\text{begin}$, $q_\text{mid}$, and $q_\text{end}$---where the same document appears in all three, each time paired with a different position-targeted query. Details of the generation prompts are provided in Appendix~\ref{app:query-generation}.

\subsubsection{Multi-Reranker Position Verification}
\label{sec:position-verification}

The generation prompt asks the LLM to produce a query answerable from the intended target segment, but this constraint is not guaranteed: a generated query may also be answerable from a non-target segment or from multiple segments. To filter such cases, we verify each generated candidate with a panel of three cross-encoder rerankers: \textit{bge-reranker-v2-m3} \citep{bge-m3}, \textit{gte-multilingual-reranker-base} \citep{gte}, and \textit{jina-reranker-v2-base-multilingual}\footnote{\url{https://huggingface.co/jinaai/jina-reranker-v2-base-multilingual}}.

We use cross-encoder rerankers, rather than dense retrievers, as verifiers because full-interaction rerankers have been shown to be more robust to evidence position than dense embedding models \citep{empirical-study}. This reduces the risk that the filtering step itself inherits the position bias that we aim to study.

The verification rule requires unanimous agreement across rerankers. Let \(q\) be a generated query for document \(d\), and let \(t\in\mathcal{P}\) be its intended target position, where \(\mathcal{P}=\{\mathrm{begin}, \mathrm{middle}, \mathrm{end}\}\). We denote the segment at position \(i\) by \(s_i\).

For each reranker \(R\in\mathcal{R}\), we score each segment as
\begin{equation}
    r_{R,i} = R(q, s_i), \quad i\in\mathcal{P}.
\end{equation}
The candidate is retained only if every reranker scores the target segment at least \(\delta\) higher than the strongest non-target segment:
\begin{equation}
    r_{R,t} - \max_{u\neq t} r_{R,u} \geq \delta,
    \quad \forall R\in\mathcal{R}.
\end{equation}
The maximum is taken over the two non-target positions. Thus, even the least favorable reranker must prefer the intended target segment by at least the margin threshold \(\delta\).

All main experiments use a margin threshold of \(\delta=0.3\). Appendix~\ref{app:filtering} reports filtering statistics under different margin thresholds and an independent segment-wise LLM audit. We refer to the candidates that pass this rule as the retained pool.

\subsubsection{Controlled Training Set Sampling}
\label{sec:controlled-sampling}

Applying the multi-reranker position-verification rule with margin threshold \(\delta=0.3\) yields 481,236 retained candidate examples for training. Table~\ref{tab:retained-pool} reports the retained pool by length bin and target position.

The retained pool is not position-balanced, so we do not train on it directly. Instead, we construct the final training sets by downsampling within length-position cells. The smallest retained length-position cell is the middle-position cell in the 4096--8192 length bin, which contains 8,189 examples. This cell determines the sampling budget for the controlled training configurations defined in Section~\ref{sec:experiment-design}.

This downsampling step ensures that the final training sets use the same number of examples from each length bin, rather than inheriting the uneven length and position counts of the retained pool. As a result, later comparisons are not driven by differences in training size or document length.\begin{table}[t]
\centering
\small
\begin{tabular}{lrrr}
\toprule
\textbf{Length Bin} & \textbf{Begin} & \textbf{Middle} & \textbf{End} \\
\midrule
256--512    & 105{,}652 & 13{,}934 & 21{,}405 \\
512--1024   &  86{,}495 & 16{,}660 & 21{,}427 \\
1024--2048  &  60{,}357 & 13{,}594 & 16{,}691 \\
2048--4096  &  43{,}946 & 10{,}527 & 13{,}363 \\
4096--8192  &  39{,}200 &  \textbf{8{,}189} &  9{,}796 \\
\midrule
Total       & 335{,}650 & 62{,}904 & 82{,}682 \\
\bottomrule
\end{tabular}
\caption{Retained candidate examples by length bin and target position after the multi-reranker position-verification rule with margin threshold \(\delta=0.3\), before downsampling.}
\vspace{-0.7em}
\label{tab:retained-pool}
\end{table}

\subsection{Position-Controlled Experiment Design}
\label{sec:experiment-design}

Our experimental design tests whether retrieval-level position bias follows the positional distribution of training data across models with different architectural properties.

\subsubsection{Model Selection and Initial Tendencies}
\label{sec:blank-slate}

We select eight pretrained models without retrieval-specific fine-tuning, spanning encoder and decoder architectures, multiple positional encodings, and different pooling strategies. This diversity is central to our design: if models with fundamentally different positional processing develop bias patterns that mirror their training data, the bias cannot be attributed to any single architectural property.

Before retrieval fine-tuning, these models are not perfectly position-neutral at the representation level: encoder models show mild primacy tendencies, while decoder models show recency tendencies (Appendix~\ref{app:pretrained-tendencies}). This makes the test stricter: a data-driven effect should appear despite different initial tendencies, and in some configurations must reverse them.

\subsubsection{Controlled Training Configurations}
\label{sec:training-configurations}

Each model is fine-tuned as a dense retriever under four configurations that differ only in the target-position distribution of training queries, expressed as \textit{begin:middle:end} ratios. Three concentrated configurations---100:0:0 (begin; $\mathcal{M}_B$), 0:100:0 (middle; $\mathcal{M}_M$), and 0:0:100 (end; $\mathcal{M}_E$)---restrict all queries to a single target position. The uniform configuration, 33:33:33 ($\mathcal{M}_U$), samples evenly across all three target positions.

All four configurations are sampled from the \(\delta=0.3\) retained pool using the per-bin budget defined in Section~\ref{sec:controlled-sampling}. Each concentrated configuration samples 8,189 examples from its target position in each length bin, yielding 40,945 training examples. The uniform configuration randomly samples 2,729 examples from each target position within each length bin, yielding 40,935 training examples. Thus, the configurations are matched in training scale and document-length distribution up to the integer split required by the uniform setting.

This yields 32 training runs: 8 base models \(\times\) 4 training configurations. After training, we evaluate each model on position-aware benchmarks to measure how retrieval performance varies across target positions. If bias is data-driven, concentrated configurations should favor their respective target positions, while uniform training should reduce position sensitivity.

\subsection{Mechanism and Predictions}
\label{sec:mechanism}
 
\label{sec:mechanism}
We hypothesize that the directional effects in our experiments arise
from shortcut learning \citep{shortcut-learning-in-deep-neural-networks, contrastive-learning-shortcut}. InfoNCE-style contrastive objectives reward
any feature that separates positives from negatives. When evidence
position correlates with relevance in the training data,
position-linked features become discriminative and can be acquired as
shortcuts; under a position-balanced distribution, position carries no
discriminative value, leaving content matching as the primary learning
signal. We present this as an empirically supported hypothesis about
learned features rather than a formal analysis of the optimization
dynamics.

This account makes four testable predictions. (P1)~Reversing the
position--relevance correlation reverses the bias direction
(\S\ref{sub:directional-effect};
Appendix~\ref{app:posir-mirror}). (P2)~Removing the correlation
through uniform training flattens position-wise performance while
maintaining competitive mean performance
(\S\ref{sub:psi-reduction}--\S\ref{sub:mean-performance}).
(P3)~The learned preference is visible at the representation level, in
query--document similarities and document embeddings
(\S\ref{sub:qd-similarity}--\S\ref{sub:doc-embeddings}). (P4)~When the
data signal opposes a model's pre-existing positional tendency, the
data signal dominates (Appendix~\ref{app:pretrained-tendencies}). A
formal account of how data distributions induce positional
features---the data-side counterpart to the architectural analysis of
\citet{emergence-of-position-bias}---is left to future work, which our released
position-controlled datasets are designed to support.
\section{Experimental Setups}

\label{sec:experimental-setup}

\subsection{Base Models}
\label{sec:base-models}

\begin{table}[t]
\centering
\resizebox{\columnwidth}{!}{%
\begin{tabular}{lcccccc}
\toprule
\textbf{Model} & \textbf{Type} & \textbf{PE} & \textbf{Pool} & \textbf{Params} & \textbf{Max Len} \\
\midrule
BERT-base        & Enc & APE   & CLS    & 110M  & 512 \\
Longformer-b     & Enc & APE   & Mean   & 149M  & 4k \\
ModernBERT-b     & Enc & RoPE  & CLS    & 149M  & 8k \\
ModernBERT-l     & Enc & RoPE  & CLS    & 395M  & 8k \\
\midrule
GPT-2-medium     & Dec & APE   & Last-t & 355M  & 1k \\
BLOOM-560M       & Dec & ALiBi & Last-t & 560M  & 2k \\
TinyLlama-NoPE   & Dec & NoPE  & Last-t & 1.1B  & 2k \\
Qwen3-0.6B       & Dec & RoPE  & Last-t & 0.6B  & 32k \\
\bottomrule
\end{tabular}
}
\caption{Overview of the eight pretrained models used in controlled fine-tuning. PE denotes positional encoding; Pool denotes the document-pooling strategy; Max Len denotes the maximum input length.}
\vspace{-0.7em}
\label{tab:model-overview}
\end{table}

Table~\ref{tab:model-overview} lists the eight pretrained base models and their
architectural properties. On the encoder side, we include
BERT-base~\citep{bert}, ModernBERT-base and
ModernBERT-large~\citep{modern-bert}, and
Longformer-base~\citep{longformer}; on the decoder side,
GPT-2-medium~\citep{gpt2}, BLOOM-560M~\citep{bloom},
TinyLlama-NoPE~\citep{tinyllama}, and Qwen3-0.6B~\citep{qwen3}.
ModernBERT-base and large share the same architecture
at different scales, enabling a within-architecture scale comparison.
TinyLlama-NoPE, which lacks explicit positional encoding, tests
whether explicit positional encoding is a necessary condition for bias emergence.

\subsection{Training Details}
\label{sec:training-details}

All eight models are fine-tuned as bi-encoder
retrievers using InfoNCE loss with chunk-aware negatives: each
batch is drawn from a single length bin so that all negatives
share the same document length as the positive. We avoid hard
negative mining, as mining strategies may introduce
position-dependent confounds. All hyperparameters are held
constant across the four configurations within each model; the
only variable is the positional distribution of training data.
Full training details are provided in
Appendix~\ref{app:training-details}.

\subsection{Evaluation}
\label{sec:evaluation}

We evaluate all trained models on three position-aware benchmarks: \textsc{SQuAD-PosQ}, \textsc{FineWeb-PosQ}~\citep{empirical-study}, and \textsc{PosIR}~\citep{posir}. Since \textsc{FineWeb-PosQ} and \textsc{PosIR} contain longer passages, we evaluate these benchmarks only on models with sufficient context length: ModernBERT-base, ModernBERT-large, and Qwen3-0.6B. We additionally evaluate on three BEIR datasets~\citep{beir}---SciFact, HotpotQA, and FEVER---where the provided annotations allow us to identify the position of evidence, enabling analysis of how the training distributions affect performance under standard retrieval settings.

We report nDCG@10 computed separately for each positional bin of the
benchmark. To summarize position sensitivity as
a single scalar, we adopt the Position Sensitivity Index (PSI)
proposed by \citet{empirical-study}:
\begin{equation}
    \text{PSI} = 1 - \frac{\min(s)}{\max(s)}, \quad
    \text{where } \max(s) > 0
\end{equation}
and $s$ collects the metric scores across the positional bins of each benchmark.
A PSI of 0 indicates
perfect positional robustness; higher values indicate greater
sensitivity.

As characterized by \citet{empirical-study}, PSI takes a worst-case
perspective: unlike dispersion measures such as the standard deviation
or the coefficient of variation \citep{arachchige2022robust}, it
quantifies the maximum relative degradation across positional buckets.
This suits our setting, where the practical concern is systematic
failure at specific evidence positions rather than overall score
variability. Because PSI is scale-invariant, we interpret it alongside
mean performance to ensure that low PSI does not merely reflect
uniformly poor retrieval.
\section{Experimental Results}
\label{sec:results}

% Usage:
% 1. Upload `main_results.pdf` to your Overleaf project root (or adjust the path below).
% 2. Ensure you have `\usepackage{graphicx}` in the preamble.
% 3. `\input{overleaf_main_results_figure.tex}` where you want the figure to appear.

\begin{figure*}[t]
    \centering
    \includegraphics[width=\textwidth]{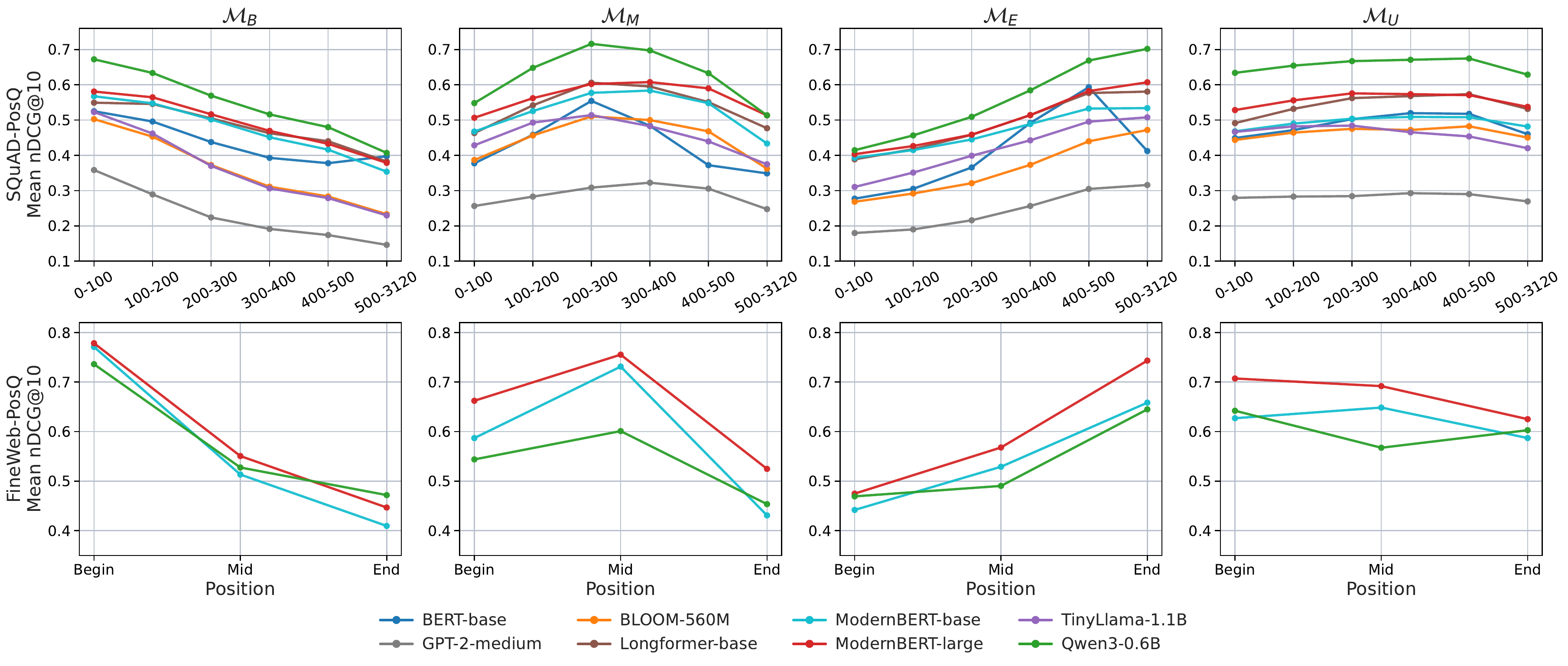}
    \vspace{-2.0em}
    \caption{Position-wise nDCG@10 across training configurations. The top row reports SQuAD-PosQ, and the bottom row reports FineWeb-PosQ. Columns correspond to configurations, $\mathcal{M}_B$, $\mathcal{M}_M$, $\mathcal{M}_E$, and $\mathcal{M}_U$; lines denote evaluated base models.}
    \label{fig:main-results}
\vspace{-0.5em}
\end{figure*}

\begin{table*}[t]
\centering
\small
\begin{tabular}{l cccc cccc}
\toprule
& \multicolumn{4}{c}{\textbf{NDCG@10}} & \multicolumn{4}{c}{\textbf{PSI} $\downarrow$} \\
\cmidrule(lr){2-5} \cmidrule(lr){6-9}
\textbf{Model} & $\mathcal{M}_B$ & $\mathcal{M}_M$ & $\mathcal{M}_E$ & $\mathcal{M}_U$ & $\mathcal{M}_B$ & $\mathcal{M}_M$ & $\mathcal{M}_E$ & $\mathcal{M}_U$ \\
\midrule
\multicolumn{9}{l}{\textsc{SQuAD-PosQ}} \\[2pt]
BERT-base        & 0.438 & 0.432 & 0.408 & \textbf{0.487} & 0.281 & 0.371 & 0.533 & \textbf{0.136} \\
Longformer-base  & 0.481 & 0.539 & 0.489 & \textbf{0.543} & 0.304 & 0.236 & 0.331 & \textbf{0.143} \\
ModernBERT-base  & 0.466 & \textbf{0.520} & 0.463 & 0.516 & 0.433 & 0.286 & 0.341 & \textbf{0.088} \\
ModernBERT-large & 0.490 & \textbf{0.564} & 0.499 & 0.557 & 0.348 & 0.166 & 0.335 & \textbf{0.082} \\
GPT-2-medium     & 0.231 & \textbf{0.287} & 0.244 & 0.283 & 0.592 & 0.233 & 0.431 & \textbf{0.080} \\
BLOOM-560M       & 0.359 & 0.447 & 0.361 & \textbf{0.465} & 0.536 & 0.290 & 0.431 & \textbf{0.080} \\
TinyLlama-NoPE   & 0.362 & 0.455 & 0.418 & \textbf{0.462} & 0.561 & 0.271 & 0.389 & \textbf{0.132} \\
Qwen3-0.6B       & 0.546 & 0.626 & 0.556 & \textbf{0.655} & 0.395 & 0.283 & 0.409 & \textbf{0.068} \\
\midrule
\multicolumn{9}{l}{\textsc{FineWeb-PosQ}} \\[2pt]
ModernBERT-base  & 0.554 & 0.570 & 0.571 & \textbf{0.640} & 0.476 & 0.422 & 0.343 & \textbf{0.108} \\
ModernBERT-large & 0.592 & 0.647 & 0.595 & \textbf{0.675} & 0.426 & 0.305 & 0.361 & \textbf{0.116} \\
Qwen3-0.6B       & 0.578 & 0.533 & 0.535 & \textbf{0.604} & 0.359 & 0.245 & 0.272 & \textbf{0.116} \\
\bottomrule
\end{tabular}
\caption{Mean nDCG@10 and Position Sensitivity Index (PSI) across training configurations. The upper block reports SQuAD-PosQ for all eight models; the lower block reports FineWeb-PosQ for models with sufficient context length. Higher is better for nDCG@10; lower is better for PSI. Best values for each model and metric are in \textbf{bold}. PosIR results for the same three long-context models are in Table 12 (Appendix~\ref{app:posir-main}).}
\label{tab:main-results}
\vspace{-1.0em}
\end{table*}

\subsection{Training Distribution Sets the Bias Direction}
\label{sub:directional-effect}

Figure~\ref{fig:main-results} shows a consistent directional effect: retrieval performance peaks near the position emphasized during fine-tuning. Begin-trained retrievers ($\mathcal{M}_B$) favor early evidence, mid-trained retrievers ($\mathcal{M}_M$) favor middle evidence, and end-trained retrievers ($\mathcal{M}_E$) favor later evidence, consistently across all eight base models. In contrast, uniformly trained retrievers ($\mathcal{M}_U$) do not exhibit a comparable single-position preference; their position-wise curves are flatter, providing an initial indication that balanced training weakens the learned positional shortcut.

Representative cases illustrate the magnitude of this shift in both short- and long-passage position-aware benchmarks. On SQuAD-PosQ, Qwen3-0.6B scores 0.672 in the 0--100 position bucket under begin training but 0.415 under end training; in the 500--3120 bucket, the pattern reverses, with end training scoring 0.702 versus 0.407 for begin training. On FineWeb-PosQ, ModernBERT-large follows the same pattern: when evidence appears at the beginning, the $\mathcal{M}_B$ scores 0.778, compared with 0.475 for the $\mathcal{M}_E$; when evidence appears at the end, the $\mathcal{M}_E$ scores 0.743, compared with 0.447 for the $\mathcal{M}_B$. The pattern also appears in TinyLlama-NoPE. Since causal masking, segment boundaries, padding, and last-token pooling still provide implicit position channels, we read this as a weakest-explicit-position condition: directional bias emerges without explicit positional encoding, though not necessarily without positional signal (Appendix~\ref{app:implicit-channels}).
% The pattern also appears in TinyLlama-NoPE, indicating that explicit positional encodings are not required for retrieval-level position bias to emerge.

Overall, these results show that retrieval-level bias direction can be redirected by the positional distribution of fine-tuning data, indicating that architecture alone does not fix the observed bias direction. Appendix~\ref{app:posir-mirror} provides an additional mirror-reversal diagnostic that confirms the same directional effect under document reversal, consistent with prediction P1 (\S\ref{sec:mechanism}).

\subsection{Balanced Training Reduces Positional Sensitivity}
\label{sub:psi-reduction}

Table~\ref{tab:main-results} shows a consistent pattern on SQuAD-PosQ and FineWeb-PosQ, and Appendix~\ref{app:posir-main} reports the same pattern on PosIR: $\mathcal{M}_U$ is the least sensitive to evidence location. It achieves the lowest PSI for all eight models on SQuAD-PosQ and for all three evaluated models on FineWeb-PosQ, indicating that balanced training produces more stable retrieval performance across positions.

On SQuAD-PosQ, $\mathcal{M}_U$ reduces PSI by 57--87\% relative to the worst skewed configuration for every model. For example, GPT-2-medium drops from 0.592 under begin training to 0.080 under uniform training, Qwen3-0.6B drops from 0.409 under end training to 0.068, and Longformer-base drops from 0.331 under end training to 0.143. The same pattern holds on FineWeb-PosQ: ModernBERT-base drops from 0.476 to 0.108, ModernBERT-large from 0.426 to 0.116, and Qwen3-0.6B from 0.359 to 0.116.

These results show that position-balanced training does not merely move the bias to a different evidence position. Instead, it makes retrieval performance more consistent across positions, so the model is less affected by where the relevant evidence appears.

\subsection{Sensitivity Reduction with Competitive Mean Performance}
\label{sub:mean-performance}

Table~\ref{tab:main-results} shows that $\mathcal{M}_U$ consistently  achieves the lowest PSI across position-aware benchmarks. On SQuAD-PosQ, $\mathcal{M}_U$ achieves the highest mean nDCG@10 for five of the eight models. For the remaining three models, its gap to the best skewed configuration is marginal (0.004--0.007). The pattern is even clearer on FineWeb-PosQ, where $\mathcal{M}_U$ achieves the highest mean nDCG@10 for all three evaluated models. Thus, position-balanced training reduces sensitivity to evidence location while maintaining competitive retrieval performance in this controlled setting, as predicted by P2.

These results also clarify the limitation of skewed training. A skewed model can perform well when the evidence appears at its trained position, but this gain often comes with larger drops at other positions. In contrast, $\mathcal{M}_U$ avoids relying on a single evidence location, leading to more stable retrieval across positions with competitive retrieval performance.

\subsection{Evidence-Location Skew in Standard Benchmarks}
\label{sub:beir-skew}

After evaluating models on controlled position-aware benchmarks, we test whether training-induced positional priors also affect performance under the standard BEIR evaluation setting, where evidence location is not controlled. Figure~\ref{fig:beir-evidence-start} shows that the three BEIR subsets differ in their evidence-location distributions. HotpotQA and FEVER are strongly concentrated near the beginning, while SciFact is broader and less early-concentrated.

Table~\ref{tab:beir} aligns with these distributional patterns. Across the three BEIR subsets and all eight base models, the begin-trained model $\mathcal{M}_B$ achieves the highest average nDCG@10: 0.393, followed by $\mathcal{M}_U$ at 0.345, $\mathcal{M}_M$ at 0.241, and $\mathcal{M}_E$ at 0.220. The advantage of $\mathcal{M}_B$ over $\mathcal{M}_U$ is largest on the most early-skewed subsets, with a gap of +0.134 on FEVER and +0.054 on HotpotQA. In contrast, this advantage disappears when evidence is less concentrated near the beginning: the gap reverses on SciFact.

These results suggest that standard benchmark scores can partly reflect evidence-location skew. When evaluation data place much of the relevant evidence near the beginning, a model with an early-position prior can obtain higher scores even if it is less robust to evidence appearing elsewhere. The BEIR results therefore indicate benchmark-specific gains rather than evidence-location robustness.

\section{Analyses}
\label{sec:analyses}

\begin{table}[t]
\centering
\small
\begin{tabular}{lrrrr}
\toprule
BEIR subset & $\mathcal{M}_B$ & $\mathcal{M}_M$ & $\mathcal{M}_E$ & $\mathcal{M}_U$ \\
\midrule
SciFact & 0.351 & \underline{0.368} & 0.340 & \textbf{0.393} \\
HotpotQA & \textbf{0.338} & 0.192 & 0.165 & \underline{0.284} \\
FEVER & \textbf{0.491} & 0.164 & 0.156 & \underline{0.357} \\
\midrule
Average & \textbf{0.393} & 0.241 & 0.220 & \underline{0.345} \\
\bottomrule
\end{tabular}
\caption{BEIR nDCG@10 averaged over all eight models. Best values are in \textbf{bold}; second-best values are \underline{underlined}. Full results are reported in Table~\ref{tab:beir-full-results}.}
\label{tab:beir}
\end{table}

\begin{figure*}[t]
\centering
\includegraphics[width=\textwidth]{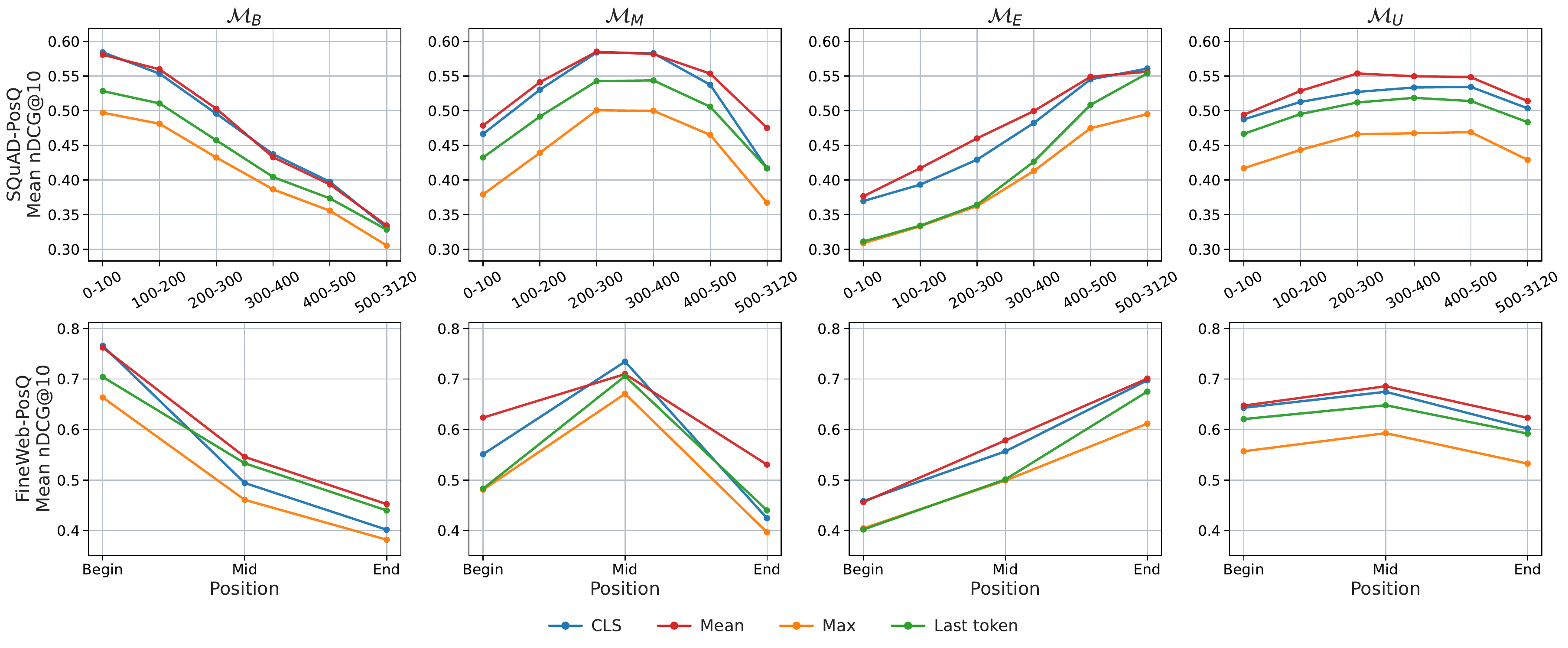}
\vspace{-2.0em}
\caption{Position-wise nDCG@10 for ModernBERT-base under four pooling strategies. The top and bottom rows report SQuAD-PosQ and FineWeb-PosQ, respectively. Columns correspond to begin-, middle-, end-, and uniform-trained retrievers; lines denote pooling strategies.}
\vspace{-1.0em}
\label{fig:pooling}
\end{figure*}

\begin{table}[t]
\centering
\small
\resizebox{\columnwidth}{!}{%
\begin{tabular}{lcccc}
\toprule
\textbf{Model} & \textbf{Config} & \textbf{Peak} & \textbf{Lowest} & \textbf{$\Delta$} \\
\midrule
\multirow{4}{*}{ModernBERT-base}
  & $\mathcal{M}_B$ & 1 & 9 & 21.5 \\
  & $\mathcal{M}_M$ & 4 & 10 & 9.4 \\
  & $\mathcal{M}_E$ & 9 & 1 & 20.6 \\
  & $\mathcal{M}_U$ & 10 & 2 & 1.9 \\
\midrule

\multirow{4}{*}{Qwen3-0.6B}
  & $\mathcal{M}_B$ & 1 & 10 & 21.5 \\
  & $\mathcal{M}_M$ & 5 & 10 & 27.1 \\
  & $\mathcal{M}_E$ & 9 & 1 & 20.6 \\
  & $\mathcal{M}_U$ & 9 & 10 & 5.5 \\
\bottomrule
\end{tabular}
}
\caption{Evidence-moving cosine analysis, where Peak and Lowest denote the insertion positions with the highest and lowest query-document cosine similarity, and $\Delta$ is the peak-minus-lowest cosine difference multiplied by \(10^3\). Full results are reported in Table~\ref{tab:evidence-moving-full}.}
\label{tab:evidence-moving}
\end{table}
\begin{figure*}[t]
    \centering
    \includegraphics[width=\textwidth]{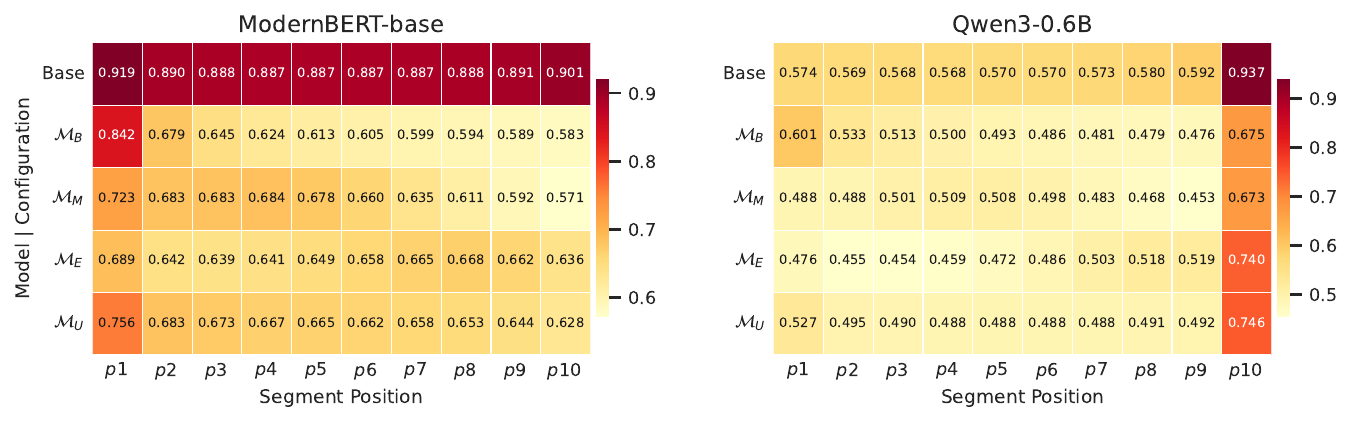}
    \vspace{-2.0em}
    \caption{
        Mean cosine similarity between full-document embeddings and segment embeddings ($p_1$--$p_{10}$) for ModernBERT-base and Qwen3-0.6B. Each column corresponds to one of ten equal-length document segments. Full results for all eight models in Figure~\ref{fig:doc-segment-all}.
    }
    \vspace{-1.5em}
    \label{fig:doc-segment-modernbert-qwen}
\end{figure*}

\subsection{Query--Document Similarity Follows the Training Distribution}
\label{sub:qd-similarity}

% The ranking-level bias observed in Section~\ref{sec:results} could
% arise from changes in the document representation itself or from
% other factors in the retrieval pipeline. To test whether the
% document embedding geometry directly encodes positional preference,
% we adapt the evidence-moving experiment of \citet{dwell}: for each
% query-document pair, the query-relevant evidence is reinserted at
% 10 uniformly spaced positions within the same document while all
% other content is held fixed, and query-document cosine similarity
% is measured at each position.

% Table~\ref{tab:evidence-moving} reports the results. For both
% ModernBERT-base and Qwen3-0.6B, peak similarity shifts to match
% the training distribution: begin-trained models peak at position~1, mid-trained at position~5, and end-trained at position~8. Uniform training compresses the cosine range from 17.8--24.8 under skewed configurations to 2.7--5.2, indicating that balanced training reduces positional preference at the representation level. Full results for all eight models are provided in Appendix~\ref{app:passage-moving}.

We first test whether the ranking-level bias observed in Section~\ref{sec:results} is reflected in query-document similarity. Following \citet{dwell}, we use MS MARCO query-document pairs and relocate the query-relevant evidence to ten uniformly spaced positions within the same document while keeping the remaining content fixed.

Table~\ref{tab:evidence-moving} shows that the highest-similarity position follows the fine-tuning distribution. For both ModernBERT-base and Qwen3-0.6B, $\mathcal{M}_B$ models peak at position~1 and $\mathcal{M}_E$ models peak at position~9. $\mathcal{M}_M$ models peak in the middle range, at p4 for ModernBERT-base and p5 for Qwen3-0.6B. Uniform training produces a flatter pattern: the peak-to-lowest gap drops to 1.9 for ModernBERT-base and 5.5 for Qwen3-0.6B, much smaller than under the concentrated settings.

These results suggest that positional preference is reflected in query-document embedding similarities, consistent with prediction P3. Fine-tuning changes which evidence locations appear most similar to the query, rather than affecting only the final ranking output. Full results for all eight models are provided in Appendix~\ref{app:passage-moving}.

% The ranking-level bias observed in Section~\ref{sec:results} could arise from the geometry of the learned embedding space or from other factors in the retrieval pipeline. To test whether query-document similarity itself encodes positional preference, we adapt the evidence-moving setup of \citet{dwell}. For each query-document pair, we insert the query-relevant evidence at ten uniformly spaced positions within the same document, keep all other content fixed, and measure query-document cosine similarity at each position.

% Table~\ref{tab:evidence-moving} shows that the highest-similarity position follows the training distribution. For both ModernBERT-base and Qwen3-0.6B, the begin-trained model peaks at position~1, the mid-trained model peaks at position~5, and the end-trained model peaks at position~9. Uniform training substantially reduces the peak-to-lowest similarity gap: for ModernBERT-base, the range drops to 2.7 under uniform training, compared with 21.2 under begin training and 18.5 under end training; for Qwen3-0.6B, it drops to 5.2, compared with 21.5, 24.8, and 17.8 under the concentrated configurations.

% These results indicate that the learned embedding space directly reflects the positional distribution of the fine-tuning data. The model is not only biased at the final ranking stage; fine-tuning changes which evidence locations appear most similar to the query. Full results for all eight models are provided in Appendix~\ref{app:passage-moving}.

% \subsection{How does training reshape document representations?}
\subsection{Fine-Tuning Reshapes Document Representations}
\label{sub:doc-embeddings}

We next examine whether the document embedding itself reflects positional preference, independent of any query. For each document, we measure cosine similarity between the full-document embedding and embeddings of its ten equal-length segments. Because this analysis is query-free, it can be applied both before and after retrieval fine-tuning. Figure~\ref{fig:doc-segment-modernbert-qwen} shows ModernBERT-base and Qwen3-0.6B; full results are in Figure~\ref{fig:doc-segment-all}.

Before retrieval fine-tuning, the pretrained base models show only mild initial positional tendencies: ModernBERT-base is slightly closer to early segments, while Qwen3-0.6B is nearly flat across segments~1--9, with a final-segment spike likely caused by last-token pooling. After fine-tuning, however, the similarity profiles shift toward the training distribution. In Qwen3-0.6B, begin training raises similarity to segment~1 while lowering later non-final segments, end training reverses this pattern, and uniform training compresses the profile across positions. ModernBERT-base shows the same qualitative trend.

These results show that retrieval fine-tuning can reshape document representations, not only query-document matching scores. Although the base checkpoints may retain weak positional tendencies, retrieval fine-tuning largely redirects them toward the positional distribution of the training data.

% \subsection{Is position bias invariant to pooling strategy?}
\subsection{Robustness of the Directional Effect to Pooling Strategy}
\label{sub:pooling}

Finally, we examine whether the observed bias direction depends on the pooling strategy. We train ModernBERT-base under the same four positional training distributions using CLS, mean, max, and last-token pooling.

Figure~\ref{fig:pooling} shows that, in this ModernBERT-base ablation, pooling changes absolute retrieval performance but the observed retrieval-level preference remains aligned with the fine-tuning position distribution. Across the four pooling choices tested, position-skewed training leads models to favor the corresponding positions on both SQuAD-PosQ and FineWeb-PosQ. Uniform training again yields a more position-balanced pattern.

These results suggest that the directional effect is not an artifact of a single pooling choice within ModernBERT-base. In this controlled ablation, changing the fine-tuning position distribution has a larger observed effect on retrieval-level bias direction than changing the pooling method.
\section{Conclusion}
We fine-tuned eight architecturally diverse dense retrievers on position-controlled data and found that the direction of retrieval-level position bias follows the positional distribution of the fine-tuning data rather than being fixed by the architecture. Representation analyses further suggest that fine-tuning reshapes query--document similarities and document embeddings toward the emphasized positions, while model-specific tendencies remain. Position-balanced training reduces positional sensitivity by 56--87\% on position-aware benchmarks while maintaining competitive mean retrieval performance in our controlled setting. These results identify training-position distribution as a major controllable factor in retrieval-level position bias and support balanced data curation as a practical mitigation.
\section*{Limitations}
This study uses synthetic position-targeted fine-tuning data built from
English Wikipedia with LLM-generated queries. The segment-permutation
intervention (Appendix~\ref{app:permutation}) shows that the learned
preference follows physical position rather than segment content,
although generated queries may exhibit position-correlated stylistic
variation and position remains partially entangled with discourse role
in some configurations; our conclusions therefore identify the
training-position distribution as the dominant, not the sole, factor.
The retained pool is verified by multiple rerankers and a held-out LLM
audit (Appendix~\ref{app:filtering}) rather than by human annotation.

To keep the intervention controlled, all runs share fixed
hyperparameters without hard-negative mining or early stopping;
three-seed replications on two backbones
(Appendix~\ref{app:seed-replication}) show that the directional effects
and PSI reductions are stable under data re-sampling.
Our evaluation targets retrieval-level behavior together with a controlled
RAG pipeline (Appendix~\ref{app:rag}); validating the effect in
production-scale systems, hard-negative training pipelines, and
naturally occurring long-document collections remains future work.
\section*{Ethics Statement}
This work constructs synthetic position-targeted retrieval data from English Wikipedia and LLM-generated queries. We do not collect private user data, conduct human-subject experiments, or infer protected attributes. However, because Wikipedia contains articles about real individuals, organizations, and sensitive topics, derived examples may include public names or sensitive or offensive content from the source corpus. The artifacts are intended for research on retrieval robustness and position-aware evaluation, not for deployment decisions about individuals or groups. Any released data or code will document the source data, licenses or terms of use, intended use, and known limitations.

\bibliography{references}

\clearpage

\appendix

%%%%%%%%%%%%%%%%%%%%%%%%%%%%%%%%%%%%%%%%%%%%%%%%%%%%%%%%%%%%%%%%%%%%%%%%%%%%%%%%%%%%%%%%%%%%%%%%%%%%%%%%%%%%%%%%%%%%%%

\section{Qwen3-Embedding Style Query Generation}
\label{app:query-generation}

We adopt a two-stage generation pipeline following \citet{qwen3-embedding}. Stage~1 selects a shared configuration (persona, difficulty, query length) for each document, and Stage~2 generates position-conditioned query--answer pairs using that configuration. For each document, the persona candidate set is retrieved from PersonaHub \citep{personahub} by embedding similarity using BGE-M3 (top-$k{=}20$). Both stages use GPT-4o-mini with temperature $T{=}1.0$ and top-$p{=}1.0$.

\subsection{Prompt for Configuration Selection (Stage~1)}
\label{app:prompt-stage1}

Given a document and a set of candidate personas, the model selects the most appropriate generation configuration (Figure~\ref{fig:prompt-stage1}). This configuration is shared across all three positional queries for the same document to keep persona, difficulty, and query length fixed across target positions. Here, \texttt{\{CHARACTERS\}} contains the 20 retrieved personas, each on a separate line. If any field fails to parse, the document is excluded from downstream processing.

\begin{figure}[t]
\begin{tcblisting}{promptdisplay,
  title={Stage 1 user prompt: configuration selection}}
[USER]
<task>
You are given a passage. Select the most appropriate configuration for generating a search query about this passage.
</task>
<passage> {PASSAGE} </passage>
<instructions>
Based on the passage content, select:
1. Character: A persona who would naturally search for this information
2. Difficulty: The education level appropriate for understanding this content
3. Query_Length: The appropriate length for the query
</instructions>
<options>
Character Candidates: {CHARACTERS}
Difficulties: high_school, university, phd
Query_Lengths: short (under 10 words), medium (10--20 words), long (over 20 words)
</options>
Output as JSON: {"Character": "selected character description", "Difficulty": "selected difficulty", "Query_Length": "selected length"}
\end{tcblisting}
\vspace{-0.5em}
\caption{Stage~1 prompt for configuration selection. A single configuration is determined per document and reused across all positional queries.}
\label{fig:prompt-stage1}
\vspace{-1.5em}
\end{figure}

\subsection{Prompt for Position-Conditioned Query Generation (Stage~2)}
\label{app:prompt-stage2}

For each document--position pair, the model generates a query--answer pair using the shared configuration from Stage~1 (Figure~\ref{fig:prompt-stage2}). The model receives both the full document and the target segment, and is instructed to produce a query answerable only from the target segment. The positional constraint is enforced at the prompt level but is not guaranteed by the generator. Generated pairs are subsequently validated through the multi-reranker filtering pipeline described in Section~\ref{sec:position-verification}.

\begin{figure*}[t]
\begin{tcblisting}{promptdisplay,
  title={Stage 2 user prompt: position-conditioned query generation}}
[USER]
<task>
You are {CHARACTER}. Generate a search query to find information about the TARGET SEGMENT within the document. The query should be answerable ONLY by the target segment, not by other parts of the document.
</task>
<full_document> {FULL_DOCUMENT} </full_document>
<target_segment position="{POSITION}"> {TARGET_SEGMENT} </target_segment>
<configuration>
- Your persona: {CHARACTER}
- Difficulty level: {DIFFICULTY}
- Target query length: {QUERY_LENGTH}
</configuration>
<requirements>
- Generate a query from your persona's perspective
- The query MUST require information specifically from the target segment to answer
- The query should NOT be answerable using only the other parts of the document
- Match the specified difficulty and length
- The answer must be directly extractable from the target segment
</requirements>
Output as JSON: {"query": "the generated search query", "answer": "the answer extracted from target segment", "reasoning": "brief explanation of why this query targets the specific segment"}
\end{tcblisting}
\vspace{-1.0em}
\caption{Stage~2 prompt for position-conditioned query generation. \texttt{\{POSITION\}} $\in$ \{beginning, middle, end\}. The configuration fields are inherited from Stage~1.}
\label{fig:prompt-stage2}
\vspace{-1.5em}
\end{figure*}

%%%%%%%%%%%%%%%%%%%%%%%%%%%%%%%%%%%%%%%%%%%%%%%%%%%%%%%%%%%%%%%%%%%%%%%%%%%%%%%%%%%%%%%%%%%%%%%%%%%%%%%%%%%%%%%%%%%%%%
%%%%%%%%%%%%%%%%%%%%%%%%%%%%%%%%%%%%%%%%%%%%%%%%%%%%%%%%%%%%%%%%%%%%%%%%%%%%%%%%%%%%%%%%%%%%%%%%%%%%%%%%%%%%%%%%%%%%%%
%%%%%%%%%%%%%%%%%%%%%%%%%%%%%%%%%%%%%%%%%%%%%%%%%%%%%%%%%%%%%%%%%%%%%%%%%%%%%%%%%%%%%%%%%%%%%%%%%%%%%%%%%%%%%%%%%%%%%%

\section{Reranker Filtering and Data Quality Audit}
\label{app:filtering}

This appendix provides additional validation for the multi-reranker filtering step used to construct the position-controlled training data. The goal of this filtering step is to obtain a high-confidence retained pool of candidate examples whose generated queries are grounded in the intended target segment. The retained pool is not used directly as the final training distribution; instead, final training sets are constructed by downsampling within length-position cells, as described in Section~\ref{sec:controlled-sampling}.

We first define the consensus margin used for threshold analysis, then report filtering statistics under different margin thresholds. We next validate the retained candidates with an independent segment-wise LLM audit. Finally, we summarize how the final \(\delta=0.3\) training sets are sampled from the retained pool.

%%%%%%%%%%%%%%%%%%%%%%%%%%%%%%%%%%%%%%%%%%%%%%%%%%%%%%%%%%%%%%%%%%%%%%%%%%%%%%%%%%%%%%%%%%%%%%%%%%%%%%%%%%%%%%%%%%%%%%

\subsection{Consensus Margin for Threshold Analysis}
\label{app:consensus-margin}

Section~\ref{sec:position-verification} defines the multi-reranker position verification using a margin threshold \(\delta\). For threshold analysis, we summarize the same filtering rule with a scalar consensus margin. For a candidate example with query \(q\), document \(d\), and intended target position \(t\), we define
\[
m_{\mathrm{cons}}(q,d,t)
=
\min_{R\in\mathcal{R}}
\left[
R(q,s_t)
-
\max_{u\ne t}R(q,s_u)
\right],
\]
where \(\mathcal{R}\) is the set of three rerankers and \(u\ne t\) ranges over the two non-target positions.

The consensus margin is the smallest target-vs-non-target score gap among the three rerankers. A candidate passes margin threshold \(\delta\) if and only if \(m_{\mathrm{cons}}(q,d,t)\ge\delta\). Thus, \(m_{\mathrm{cons}}\ge0\) means that all three rerankers rank the target segment above both non-target segments, while larger thresholds require stronger agreement that the query is grounded in the intended target segment.

We refer to candidates with \(m_{\mathrm{cons}}\ge0\) as non-failing candidates. The final training sets use the retained pool obtained with margin threshold \(\delta=0.3\).

%%%%%%%%%%%%%%%%%%%%%%%%%%%%%%%%%%%%%%%%%%%%%%%%%%%%%%%%%%%%%%%%%%%%%%%%%%%%%%%%%%%%%%%%%%%%%%%%%%%%%%%%%%%%%%%%%%%%%%

\subsection{Filtering Statistics and Retained-Pool Skew}
% \subsection{Filtering Statistics under Different Margin Thresholds}
\begin{table*}[t]
\centering
\small
\setlength{\tabcolsep}{5pt}
\begin{tabular}{lrrcccccc}
\toprule
& & & \multicolumn{6}{c}{Target position within retained pool} \\
\cmidrule(lr){4-9}
& &
& \multicolumn{2}{c}{Begin}
& \multicolumn{2}{c}{Middle}
& \multicolumn{2}{c}{End} \\
\cmidrule(lr){4-5}
\cmidrule(lr){6-7}
\cmidrule(lr){8-9}
Filter & Retained & \% generated & \(N\) & Share & \(N\) & Share & \(N\) & Share \\
\midrule
All generated candidates
& 2,948,006 & 100.00
& -- & -- & -- & -- & -- & -- \\

Failed top-rank check
& 945,845 & 32.08
& -- & -- & -- & -- & -- & -- \\
\midrule
\(m_{\mathrm{cons}}\ge0\)
& 2,002,161 & 67.92
& 925,013 & 46.20
& 523,440 & 26.14
& 553,708 & 27.66 \\

\(m_{\mathrm{cons}}\ge0.1\)
& 1,385,322 & 46.99
& 754,366 & 54.45
& 295,120 & 21.30
& 335,836 & 24.24 \\

\(m_{\mathrm{cons}}\ge0.2\)
& 869,738 & 29.50
& 539,882 & 62.07
& 149,127 & 17.15
& 180,729 & 20.78 \\

\(m_{\mathrm{cons}}\ge0.3\)
& 481,236 & 16.32
& 335,650 & 69.75
& 62,904 & 13.07
& 82,682 & 17.18 \\
\bottomrule
\end{tabular}
\vspace{-0.5em}
\caption{Cumulative multi-reranker filtering statistics for training candidates. Position columns report the number of retained candidates and their share within each thresholded pool. Increasing the margin threshold reduces coverage but requires stronger agreement that the generated query is grounded in the intended target segment.}
\vspace{-1.0em}
\label{tab:filtering-funnel}
\end{table*}

\label{app:filtering-statistics}

We first quantify how many generated candidates remain under different cumulative margin thresholds. Table~\ref{tab:filtering-funnel} reports the number of retained candidates, their fraction among all generated candidates, and the target-position composition of the retained pool.

The final threshold \(\delta=0.3\) retains 481,236 candidates, including 62,904 middle-targeted and 82,682 end-targeted candidates. Although stricter thresholds make the retained pool increasingly begin-skewed, the \(\delta=0.3\) pool still contains enough examples in every target position for controlled sampling.

% Retention also varies by document length. Figure~\ref{fig:length-retention-heatmap} reports the retention rate within each length bin under cumulative margin thresholds. Stricter filtering disproportionately reduces longer-document candidates, which motivates the length-position controlled sampling described in Section~\ref{sec:controlled-sampling}.

Together, these statistics show that the \(\delta=0.3\) retained pool is conservative under our model-based verification criteria and higher-confidence than lower-margin strata, but not position-balanced or length-neutral. We therefore use it only as a source pool for controlled sampling, rather than as the final training distribution.
%%%%%%%%%%%%%%%%%%%%%%%%%%%%%%%%%%%%%%%%%%%%%%%%%%%%%%%%%%%%%%%%%%%%%%%%%%%%%%%%%%%%%%%%%%%%%%%%%%%%%%%%%%%%%%%%%%%%%%

\subsection{Segment-Wise LLM Audit}
\label{app:llm-audit}

To independently validate that the reranker margin corresponds to segment-exclusive answerability, we conduct a held-out LLM audit. The audit uses a single binary judge prompt: for each candidate example, we pair the query independently with each of the three document segments and ask whether that segment contains the answer. The intended target position is not revealed to the judge. This audit is used only for post-hoc validation and is not part of the training-set construction pipeline.

Figure~\ref{fig:prompt-llm-audit} shows the exact binary prompt used for all reported audit results. The reported quantities---TargetYes, DistractorNo, and Exclusive---are aggregate metrics computed from independent segment-level yes/no judgments, not separate prompts.

\begin{figure}[t]
\begin{tcblisting}{promptdisplay,title={Segment-level LLM audit prompt}}
[SYSTEM]
You are a precise reading comprehension evaluator. Given a question and a text segment, determine whether the segment contains the answer to the question.

[USER]
Question: {question}

Text:
{segment_text}

Does the text above contain the answer to the question?
Respond in JSON: {"reasoning": "...", "answer": "yes|no"}
\end{tcblisting}
\caption{Binary segment-level prompt used for the LLM audit. Each query--segment pair is evaluated independently, and the audit metrics are computed by aggregating the resulting yes/no judgments across the target and non-target segments.}
\label{fig:prompt-llm-audit}
\end{figure}

Let \(\mathcal{P}=\{\mathrm{begin},\mathrm{middle},\mathrm{end}\}\). 
For candidate \(j\), let \(t_j\in\mathcal{P}\) be the labeled target position. 
We let \(J_{j,i}\in\{0,1\}\) denote the binary LLM judgment for whether segment 
\(s_{j,i}\) contains the answer to query \(q_j\), where \(J_{j,i}=1\) indicates 
\texttt{yes} and \(J_{j,i}=0\) indicates \texttt{no}. We also let 
\(\bar{J}_{j,i}=1-J_{j,i}\), so \(\bar{J}_{j,i}=1\) means that segment \(s_{j,i}\) 
is judged answer-absent. Let \(N\) be the number of audited candidate examples.

We report three aggregate metrics:
\[
\begin{aligned}
\mathrm{TargetYes}
&=
\frac{1}{N}
\sum_{j=1}^{N}
J_{j,t_j},
\\
\mathrm{DistractorNo}
&=
\frac{1}{2N}
\sum_{j=1}^{N}
\sum_{i\ne t_j}
\bar{J}_{j,i},
\\
\mathrm{Exclusive}
&=
\frac{1}{N}
\sum_{j=1}^{N}
J_{j,t_j}
\prod_{i\ne t_j}
\bar{J}_{j,i}.
\end{aligned}
\]
Here, \(i\ne t_j\) ranges over the two non-target positions.
\(\mathrm{TargetYes}\) measures whether the labeled target segment is judged 
answer-containing. \(\mathrm{DistractorNo}\) measures whether the two non-target 
segments are judged answer-absent. \(\mathrm{Exclusive}\) is the strictest metric: 
it requires the target segment to be judged answer-containing and both non-target 
segments to be judged answer-absent for the same candidate.

Table~\ref{tab:llm-audit-threshold} reports the audit results across non-overlapping consensus-margin strata. Unlike Table~\ref{tab:filtering-funnel}, which reports cumulative retained pools, this audit uses disjoint margin ranges so that lower-margin groups are not mixed with candidates that would also pass stricter thresholds. The final \(\delta=0.3\) retained pool corresponds to the highest-margin stratum, \(m_{\mathrm{cons}}\ge0.3\).

\begin{table}[t]
\centering
\small
\resizebox{\columnwidth}{!}{%
\begin{tabular}{lccc}
\toprule
Reranker Condition & Target Yes & Distractor No & Exclusive \\
\midrule
Failed top-rank check                    & 87.7\% & 73.8\% & 51.4\% \\
\(0\:\:\,\,\le m_{\mathrm{cons}}<0.1\)          & 93.4\% & 83.5\% & 67.0\% \\
\(0.1\le m_{\mathrm{cons}}< 0.2\)        & 93.7\% & 90.2\% & 77.7\% \\
\(0.2\le m_{\mathrm{cons}}< 0.3\)        & 94.7\% & 94.4\% & 85.3\% \\
\(0.3\le m_{\mathrm{cons}}\)        & 95.4\% & 97.0\% & 90.4\% \\
\bottomrule
\end{tabular}
}
\caption{Segment-wise LLM audit across non-overlapping consensus-margin strata. Target Yes, Distractor No, and Exclusive are aggregate metrics computed from independent binary yes/no judgments. Exclusive requires the target segment to be judged as containing the answer and both non-target segments to be judged as not containing the answer. Higher values indicate better segment exclusivity.}
\vspace{-0.6em}
\label{tab:llm-audit-threshold}
\end{table}

The exclusive rate increases monotonically across higher-margin strata, indicating that the reranker margin is an effective precision-control signal. Reranker-failed candidates have the lowest exclusive rate, while the \(m_{\mathrm{cons}}\ge0.3\) group has the highest audited exclusivity. Combined with the retained-pool statistics in Table~\ref{tab:filtering-funnel}, this supports \(\delta=0.3\) as a conservative high-precision setting: it yields the strongest audited segment exclusivity while still leaving enough candidates for controlled training-set construction.

%%%%%%%%%%%%%%%%%%%%%%%%%%%%%%%%%%%%%%%%%%%%%%%%%%%%%%%%%%%%%%%%%%%%%%%%%%%%%%%%%%%%%%%%%%%%%%%%%%%%%%%%%%%%%%%%%%%%%%

\subsection{Final Sampling from the \texorpdfstring{$\delta=0.3$}{delta=0.3} Retained Pool}
\label{app:final-sampling}

\begin{table}[t]
\centering
\small
\setlength{\tabcolsep}{4pt}
\begin{tabular}{lrrrr}
\toprule
Length bin & Begin pool & Middle pool & End pool & Budget \\
\midrule
256--512    & 105,652 & 13,934 & 21,405 & 8,189 \\
512--1024   & 86,495  & 16,660 & 21,427 & 8,189 \\
1024--2048  & 60,357  & 13,594 & 16,691 & 8,189 \\
2048--4096  & 43,946  & 10,527 & 13,363 & 8,189 \\
4096--8192  & 39,200  & 8,189  & 9,796  & 8,189 \\
\midrule
Total       & 335,650 & 62,904 & 82,682 & 40,945 \\
\bottomrule
\end{tabular}
\caption{Final sampling budget from the \(\delta=0.3\) retained pool. The smallest length-position cell is the 4096--8192 middle cell with 8,189 examples, which sets the per-bin budget for concentrated configurations. The uniform configuration samples 2,729 examples from each target position within each length bin, yielding 40,935 examples in total.}
\label{tab:final-sampling-budget}
\end{table}

The \(\delta=0.3\) retained pool is used as a high-confidence source pool, not as the final training distribution. As shown above, the retained pool is begin-skewed and retention varies by length bin. To prevent these raw counts from becoming confounding factors, we construct final training sets by downsampling within length-position cells.

Table~\ref{tab:final-sampling-budget} reports the retained cell sizes and the sampling budget used for the final training configurations. The smallest retained length-position cell is the middle-position cell in the 4096--8192 length bin, which contains 8,189 examples. This cell sets the common per-bin sampling budget for concentrated configurations.

Each concentrated configuration samples 8,189 examples from its target position in each length bin, yielding 40,945 training examples. The uniform configuration samples 2,729 examples from each target position within each length bin, yielding 40,935 training examples. The slight difference in total size comes from the integer split of 8,189 examples into three target positions.

This sampling design prevents the final training sets from inheriting the uneven position and length counts of the retained pool. As a result, comparisons across training configurations are not driven by differences in training size or document-length distribution.

%%%%%%%%%%%%%%%%%%%%%%%%%%%%%%%%%%%%%%%%%%%%%%%%%%%%%%%%%%%%%%%%%%%%%%%%%%%%%%%%%%%%%%%%%%%%%%%%%%%%%%%%%%%%%%%%%%%%%%
%%%%%%%%%%%%%%%%%%%%%%%%%%%%%%%%%%%%%%%%%%%%%%%%%%%%%%%%%%%%%%%%%%%%%%%%%%%%%%%%%%%%%%%%%%%%%%%%%%%%%%%%%%%%%%%%%%%%%%
%%%%%%%%%%%%%%%%%%%%%%%%%%%%%%%%%%%%%%%%%%%%%%%%%%%%%%%%%%%%%%%%%%%%%%%%%%%%%%%%%%%%%%%%%%%%%%%%%%%%%%%%%%%%%%%%%%%%%%

\begin{figure*}[tp]
    \centering
    \includegraphics[width=\textwidth]{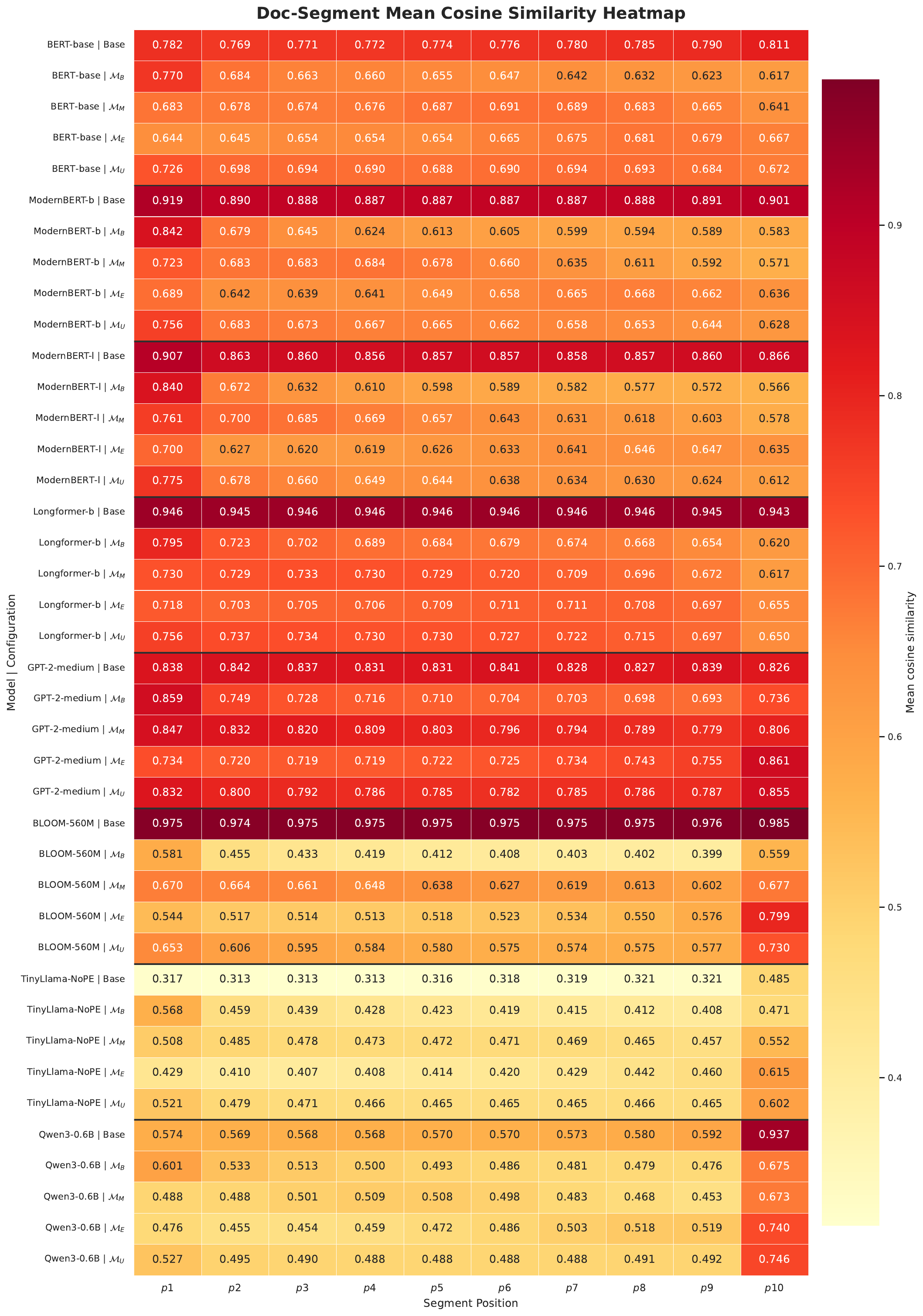}
    % \caption{
    %     Mean cosine similarity between full-document embeddings and segment embeddings ($p_1$--$p_{10}$) for all eight models. Each column corresponds to one of ten equal-length document segments.
    % }
    \caption{Full-document--segment cosine similarity for all eight models across ten equal-length document segments. ``Base'' denotes pretrained base models before retrieval fine-tuning; ``$\mathcal{M}_B$,'' ``$\mathcal{M}_M$,'' ``$\mathcal{M}_E$,'' and ``$\mathcal{M}_U$'' denote retrievers fine-tuned under the corresponding training configuration. Columns \(p_1\)--\(p_{10}\) denote segment positions.}
    \label{fig:doc-segment-all}
\end{figure*}

\section{Pre-Existing Positional Tendencies in Pretrained Models}
\label{app:pretrained-tendencies}

As noted in Section~\ref{sec:blank-slate}, pretrained models are not strictly position-neutral at the representation level. To quantify these pre-existing positional tendencies, we compute cosine similarity between each full-document embedding and the embedding of each of its ten equal-length segments. Figure~\ref{fig:doc-segment-all} (``Base'' rows, denoting pretrained models before retrieval fine-tuning) reports the results for all eight models. Here, range denotes the maximum minus minimum similarity across the ten document segments.

The base checkpoints show weak, model-specific tendencies rather than a consistent directional bias. Among encoders, Longformer-base is nearly flat (range 0.003), ModernBERT-base and ModernBERT-large show mild early preference (ranges 0.032 and 0.051), and BERT-base is shallowly U-shaped (range 0.042). Among decoders, GPT-2-medium is also nearly flat (range 0.016). BLOOM-560M, TinyLlama-NoPE, and Qwen3-0.6B show a spike at segment~10. Excluding segment~10, their profiles are almost flat, with ranges of 0.002, 0.008, and 0.024, respectively.

These initial tendencies are much smaller than the changes induced by retrieval fine-tuning. ModernBERT-base's range increases from 0.032 before fine-tuning to 0.259 under begin training, and Qwen3-0.6B's range over segments~1--9 increases from 0.024 to 0.125 under begin training and 0.065 under end training. Thus, the main experiments are conservative because the observed bias must arise despite weak, model-specific tendencies already present before fine-tuning; that the data signal overrides these tendencies is the behavior predicted by P4 (\S\ref{sec:mechanism}).

\section{Segment-Contribution Probe and Position-Channel Necessity
Control}
\label{app:implicit-channels}

Two analyses substantiate the reading of the NoPE result in
\S\ref{sub:directional-effect}: positional sensitivity predates
retrieval fine-tuning and is redirected rather than created by it, and
removing the usable position channel eliminates directional induction.

\paragraph{Leave-one-segment-out contribution.}
We measure each segment's contribution to the document embedding as
the change in the embedding when that segment is removed ($n{=}600$
documents per cell; Table~\ref{tab:loo}). Before fine-tuning,
last-token-pooled Qwen3-0.6B already weights the final segment far
more heavily than earlier ones (0.102, document-bootstrap 95\% CI
[0.094, 0.110], versus 0.010--0.014), indicating that the model's
rotary encoding, causal structure, and last-token pooling jointly
yield positional sensitivity before any retrieval training.
Fine-tuning redirects and amplifies this sensitivity rather than
creating it de novo: $\mathcal{M}_B$ reverses the pre-existing end
emphasis, moving the largest contribution to the beginning (0.257),
$\mathcal{M}_E$ reinforces and strengthens it (0.278), $\mathcal{M}_M$
shifts it to the middle (0.191, versus 0.135/0.109 for the
beginning/end), and $\mathcal{M}_U$ spreads it across segments
(0.059--0.088).

\paragraph{Position-channel necessity control.}
As a stricter test, we fine-tune BERT-base with its absolute
positional embeddings zeroed and frozen (CLS pooling) under three
configurations ($\mathcal{M}_B$, $\mathcal{M}_E$, $\mathcal{M}_U$).
The directional slope (the least-squares slope of nDCG@10 over the six
SQuAD-PosQ bin indices) collapses for the skewed arms
($\mathcal{M}_B$: $-0.0297 \to +0.0010$; $\mathcal{M}_E$: $+0.0476
\to -0.0001$), while $\mathcal{M}_U$ remains near zero as before
($+0.0060 \to +0.0014$) and retrieval itself does not collapse
(nDCG@10 remains within 0.39--0.47): removing the usable position
channel eliminates directional induction without collapsing the model.
A residual inverted-U shape shared by all three arms remains, which we
attribute to truncation or data artifacts rather than directional
induction, as it is common across configurations. This control
establishes necessity only for this bidirectional, short-document
setting. Together, the two analyses support reading the
TinyLlama-NoPE result as a weakest-explicit-position condition rather
than as evidence that positional signal is absent.

\begin{table}[t]
\centering
\small
\begin{tabular}{lccc}
\toprule
Config & Begin & Middle & End \\
\midrule
Base            & 0.014          & 0.010          & \textbf{0.102} \\
$\mathcal{M}_B$ & \textbf{0.257} & 0.016          & 0.029          \\
$\mathcal{M}_M$ & 0.135          & \textbf{0.191} & 0.109          \\
$\mathcal{M}_E$ & 0.035          & 0.066          & \textbf{0.278} \\
$\mathcal{M}_U$ & 0.080          & 0.059          & 0.088          \\
\bottomrule
\end{tabular}
\caption{Leave-one-segment-out contribution to the document embedding
for Qwen3-0.6B (last-token pooling), before (Base) and after
fine-tuning under each configuration ($n{=}600$ documents per cell).
Each entry is the mean embedding change when the corresponding segment
is removed; document-bootstrap 95\% CI half-widths are at most 0.011.
For Base and the skewed configurations, the largest contribution is in
\textbf{bold}; $\mathcal{M}_U$ has no single dominant segment
(0.059--0.088).}
\label{tab:loo}
\end{table}

%%%%%%%%%%%%%%%%%%%%%%%%%%%%%%%%%%%%%%%%%%%%%%%%%%%%%%%%%%%%%%%%%%%%%%%%%%%%%%%%%%%%%%%%%%%%%%%%%%%%%%%%%%%%%%%%%%%%%%
%%%%%%%%%%%%%%%%%%%%%%%%%%%%%%%%%%%%%%%%%%%%%%%%%%%%%%%%%%%%%%%%%%%%%%%%%%%%%%%%%%%%%%%%%%%%%%%%%%%%%%%%%%%%%%%%%%%%%%
%%%%%%%%%%%%%%%%%%%%%%%%%%%%%%%%%%%%%%%%%%%%%%%%%%%%%%%%%%%%%%%%%%%%%%%%%%%%%%%%%%%%%%%%%%%%%%%%%%%%%%%%%%%%%%%%%%%%%%

\section{Training Details}
\label{app:training-details}

\newcommand{\hf}[1]{\href{https://huggingface.co/#1}{\texttt{#1}}}

\begin{table}[t]
\centering
\resizebox{\columnwidth}{!}{%
\begin{tabular}{ll}
\toprule
Model & Checkpoint identifier \\
\midrule
BERT-base & \hf{google-bert/bert-base-uncased} \\
Longformer-base & \hf{allenai/longformer-base-4096} \\
ModernBERT-base & \hf{answerdotai/ModernBERT-base} \\
ModernBERT-large & \hf{answerdotai/ModernBERT-large} \\
GPT-2-medium & \hf{openai-community/gpt2-medium} \\
BLOOM-560M & \hf{bigscience/bloom-560m} \\
TinyLlama-NoPE-1.1B & \hf{AntNLP/TinyLlama-NoPE-1.1B} \\
Qwen3-0.6B & \hf{Qwen/Qwen3-0.6B} \\
bge-reranker-v2-m3 & \hf{BAAI/bge-reranker-v2-m3} \\
gte-multilingual-reranker-base & \hf{Alibaba-NLP/gte-multilingual-reranker-base} \\
jina-reranker-v2-base-multilingual & \hf{jinaai/jina-reranker-v2-base-multilingual} \\
\bottomrule
\end{tabular}
}
\caption{Public checkpoint identifiers for the eight base models used
in controlled fine-tuning and the three cross-encoder rerankers used
for multi-reranker position verification.}
\label{tab:checkpoint-sources}
\end{table}

\begin{table}[ht]
\centering
\small
\resizebox{\columnwidth}{!}{%
\begin{tabular}{ll}
\toprule
\textbf{Parameter} & \textbf{Value} \\
\midrule
Loss function      & InfoNCE (CachedMNRL) \\
Optimizer          & AdamW \\
Batch size         & 256 \\
Epochs             & 3 \\
Warmup ratio       & 0.1 \\
Similarity scale   & 20.0 ($\tau = 0.05$) \\
Chunk-aware negatives & Enabled \\
Hard negative mining  & None \\
Seed               & 42 \\
\bottomrule
\end{tabular}
}
\caption{Shared training hyperparameters for all 32 fine-tuning runs.}
\label{tab:shared-hyper}
\end{table}

Table~\ref{tab:checkpoint-sources} lists the public base checkpoints used in our controlled fine-tuning experiments. All 32 model configurations are trained using the Sentence Transformers library~\citep{reimers-gurevych-2019-sentence}, built on Transformers \citep{sentence-transformers}, with identical hyperparameters except for the learning rate, which varies by model scale. Table~\ref{tab:shared-hyper} lists the shared training configuration.

We set the learning rate to $4 \times 10^{-5}$ for the smaller models
(BERT-base, ModernBERT-base, Longformer-base, GPT-2-medium)
and $2 \times 10^{-5}$ for the larger models (ModernBERT-large, BLOOM-560M, TinyLlama-NoPE, Qwen3-0.6B).
All runs use the final checkpoint after three epochs with no early 
stopping or checkpoint selection.

We prepend an instruction prefix to all query inputs at both training 
and inference time: \texttt{"query: "} for encoder-only models and 
\texttt{"Retrieve a relevant passage: "} for decoder-only models. 
No prefix is applied to document inputs.

Fine-tuning all 32 model configurations took approximately 6 hours on 8 NVIDIA A100-SXM4-80GB GPUs, or about 48 GPU-hours, excluding data generation, filtering, and evaluation.

The position-controlled training sets for all four configurations ($\mathcal{M}_B$, $\mathcal{M}_M$, $\mathcal{M}_E$, $\mathcal{M}_U$; \(\delta=0.3\)) are publicly available at \url{https://huggingface.co/datasets/sionic-ai/position-bias-training-datasets}.

%%%%%%%%%%%%%%%%%%%%

\begin{figure*}[tp]
    \centering
    \includegraphics[width=\textwidth]{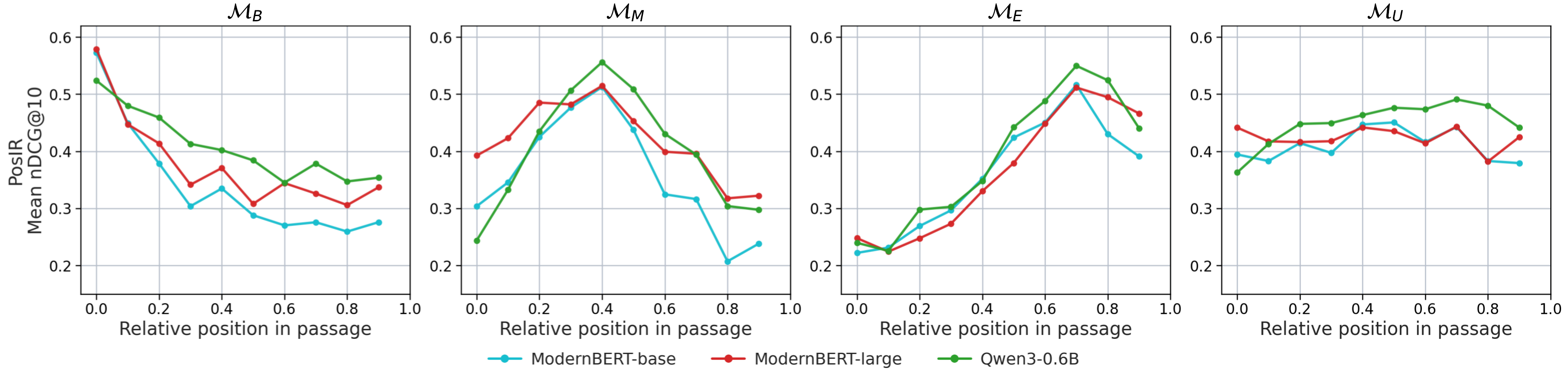}
    \vspace{-2em}
    \caption{Position-wise nDCG@10 on four selected PosIR domains: Subject Education, News Media, Law Judiciary, and Finance Economics. Columns correspond to configurations, $\mathcal{M}_B$, $\mathcal{M}_M$, $\mathcal{M}_E$, and $\mathcal{M}_U$; lines denote evaluated base models.}
    \label{fig:main-results-posir}
\end{figure*}

\begin{table}[t]
\centering
\small
\resizebox{\columnwidth}{!}{%
\begin{tabular}{l cccc cccc}
\toprule
& \multicolumn{4}{c}{\textbf{NDCG@10}} & \multicolumn{4}{c}{\textbf{PSI} $\downarrow$} \\
\cmidrule(lr){2-5} \cmidrule(lr){6-9}
\textbf{Model} & $\mathcal{M}_B$ & $\mathcal{M}_M$ & $\mathcal{M}_E$ & $\mathcal{M}_U$ & $\mathcal{M}_B$ & $\mathcal{M}_M$ & $\mathcal{M}_E$ & $\mathcal{M}_U$ \\
\midrule
% \multicolumn{9}{l}{\textsc{PosIR}} \\[2pt]
ModernBERT-b  & 0.341 & 0.359 & 0.358 & \textbf{0.411} & 0.547 & 0.596 & 0.570 & \textbf{0.158} \\
ModernBERT-l & 0.377 & 0.419 & 0.362 & \textbf{0.423} & 0.472 & 0.383 & 0.562 & \textbf{0.138} \\
Qwen3-0.6B & 0.409 & 0.401 & 0.386 & \textbf{0.450} & 0.341 & 0.562 & 0.590 & \textbf{0.261} \\
\bottomrule
\end{tabular}
}
\caption{Mean nDCG@10 and Position Sensitivity Index (PSI) on PosIR across training configurations for models with sufficient context length. Higher is better for nDCG@10; lower is better for PSI. Best values for each model and metric are in \textbf{bold}.}
\vspace{-1.0em}
\label{tab:main-results-posir}
\end{table}

\section{Additional Experimental Results: PosIR}

This appendix reports additional PosIR results: the main position-wise
results on four selected domains (\S \ref{app:posir-main}), and a
mirror-reversal diagnostic (\S \ref{app:posir-mirror}).

\subsection{Main Results on Four PosIR Domains}
\label{app:posir-main}
Figure~\ref{fig:main-results-posir} reports position-wise nDCG@10 on the PosIR four selected subset domains: Subject Education, News Media, Law Judiciary, and Finance Economics. The same directional pattern observed on SQuAD-PosQ and FineWeb-PosQ also appears on PosIR: begin-trained retrievers ($\mathcal{M}_B$) favor earlier evidence, mid-trained retrievers ($\mathcal{M}_M$) peak around the middle, and end-trained retrievers ($\mathcal{M}_E$) improve toward later evidence. Uniformly trained retrievers ($\mathcal{M}_U$) produce a flatter curve, indicating lower sensitivity to the physical location of the reference evidence.

Table~\ref{tab:main-results-posir} summarizes mean nDCG@10 and PSI. The $\mathcal{M}_U$ achieves the highest mean nDCG@10 for all three evaluated long-context models: 0.411 for ModernBERT-base, 0.423 for ModernBERT-large, and 0.450 for Qwen3-0.6B. It also achieves the lowest PSI for all three models, reducing PSI relative to the worst skewed configuration by 73.5\% for ModernBERT-base, 75.4\% for ModernBERT-large, and 55.8\% for Qwen3-0.6B.

These results show that the main position-aware findings extend to PosIR. Position-skewed training still induces position-specific preferences, while position-balanced training reduces sensitivity to evidence location with higher mean nDCG@10 on the evaluated PosIR domains.

\subsection{Mirror-Reversal Diagnostic on PosIR}
\label{app:posir-mirror}

As a counterfactual diagnostic, we divide each PosIR document into five equal contiguous segments and reverse their order ($1,2,3,4,5 \to 5,4,3,2,1$), keeping the query and relevance label fixed. Queries are stratified by the original location of their evidence: front-origin evidence (segments 1--2) moves to the back after reversal (F$\to$B), and back-origin evidence (segments 4--5) moves to the front (B$\to$F); mid-origin evidence (segment 3) is unaffected and not reported. We define the reversal front--back gap as
\begin{equation*}
\Delta_{\mathrm{rev}} = \mathrm{B{\to}F} - \mathrm{F{\to}B},
\end{equation*}
so positive values indicate better retrieval when evidence is moved to the front and negative values when it is moved to the back.

Table~\ref{tab:posir-original-reversal} reports the results. For every backbone, $\mathcal{M}_B$ has a positive $\Delta_{\mathrm{rev}}$ ($+0.194$, $+0.205$, $+0.236$) and $\mathcal{M}_E$ a negative one ($-0.126$, $-0.137$, $-0.230$): the same origin group is retrieved better once its evidence occupies the trained position, so the preference follows the current physical position, as predicted by P1 (\S\ref{sec:mechanism}). $\mathcal{M}_M$ shows an intermediate positive gap ($+0.114$ to $+0.153$), and $\mathcal{M}_U$ the smallest magnitude on every backbone ($+0.039$ to $+0.090$). Because reversal also changes discourse order, this diagnostic complements the controlled intervention in Appendix~\ref{app:permutation}.

\begin{table*}[t]
\centering
% \scriptsize
\small
\setlength{\tabcolsep}{5pt}
\begin{tabular}{lccccc}
\toprule
& \multicolumn{2}{c}{Front-origin queries} &
  \multicolumn{2}{c}{Back-origin queries} & \\
\cmidrule(lr){2-3}
\cmidrule(lr){4-5}
Model
& Orig. front & Rev. F\(\to\)B
& Orig. back & Rev. B\(\to\)F
& \(\Delta_{\mathrm{rev}}\) \\
\midrule
\multicolumn{6}{l}{\textsc{ModernBERT-base}} \\[2pt]
\(\mathcal{M}_B\)
& 0.435 & 0.187 & 0.272 & 0.382 & +0.194 \\
\(\mathcal{M}_M\)
& 0.364 & 0.238 & 0.288 & 0.375 & +0.138 \\
\(\mathcal{M}_E\)
& 0.238 & 0.458 & 0.458 & 0.331 & -0.126 \\
\(\mathcal{M}_U\)
& 0.385 & 0.353 & 0.413 & 0.413 & +0.060 \\
\midrule
\multicolumn{6}{l}{\textsc{ModernBERT-large}} \\[2pt]
\(\mathcal{M}_B\)
& 0.447 & 0.238 & 0.326 & 0.443 & +0.205 \\
\(\mathcal{M}_M\)
& 0.431 & 0.310 & 0.371 & 0.463 & +0.153 \\
\(\mathcal{M}_E\)
& 0.232 & 0.497 & 0.478 & 0.360 & -0.137 \\
\(\mathcal{M}_U\)
& 0.415 & 0.387 & 0.416 & 0.477 & +0.090 \\
\midrule
\multicolumn{6}{l}{\textsc{Qwen3-0.6B}} \\[2pt]
\(\mathcal{M}_B\)
& 0.475 & 0.353 & 0.352 & 0.589 & +0.236 \\
\(\mathcal{M}_M\)
& 0.354 & 0.417 & 0.372 & 0.531 & +0.114 \\
\(\mathcal{M}_E\)
& 0.254 & 0.680 & 0.502 & 0.450 & -0.230 \\
\(\mathcal{M}_U\)
& 0.409 & 0.581 & 0.469 & 0.620 & +0.039 \\
\bottomrule
\end{tabular}
\caption{Original and mirror-reversed PosIR performance for front- and back-origin queries. Queries are stratified by the original location of their reference evidence. Front-origin queries have evidence in segments 1--2; after reversal, this evidence moves to the back, denoted F\(\to\)B. Back-origin queries have evidence in segments 4--5; after reversal, this evidence moves to the front, denoted B\(\to\)F. \(\mathcal{M}_B\), \(\mathcal{M}_M\), and \(\mathcal{M}_E\) denote retrievers trained on begin-, middle-, and end-concentrated configurations, respectively, and \(\mathcal{{M}_U}\) denotes the uniform configuration. We define \(\Delta_{\mathrm{rev}}=\mathrm{B{\to}F}-\mathrm{F{\to}B}\), so positive values indicate a preference for evidence currently placed near the front after reversal, while negative values indicate a preference for evidence currently placed near the back.}
\label{tab:posir-original-reversal}
\end{table*}

\section{Cross-Lingual Zero-Shot Transfer}
We test whether the positional prior induced by English-only fine-tuning
transfers across languages. We use Qwen3-0.6B, the only multilingually
pretrained backbone among the three models with sufficient context
length for PosIR.

\subsection{Seven-Language Evaluation on PosIR}
\label{app:posir-multilingual}

The four retrievers
($\mathcal{M}_B$, $\mathcal{M}_M$, $\mathcal{M}_E$, $\mathcal{M}_U$),
fine-tuned exclusively on English position-controlled data, are evaluated
unchanged---without any multilingual adaptation---on the Arabic, Chinese,
French, Japanese, Korean, and Russian versions of the four PosIR domains
used in Appendix~\ref{app:posir-main}; the English results are included for
reference.

\begin{figure*}[tp]
    \centering
    \includegraphics[width=\textwidth]{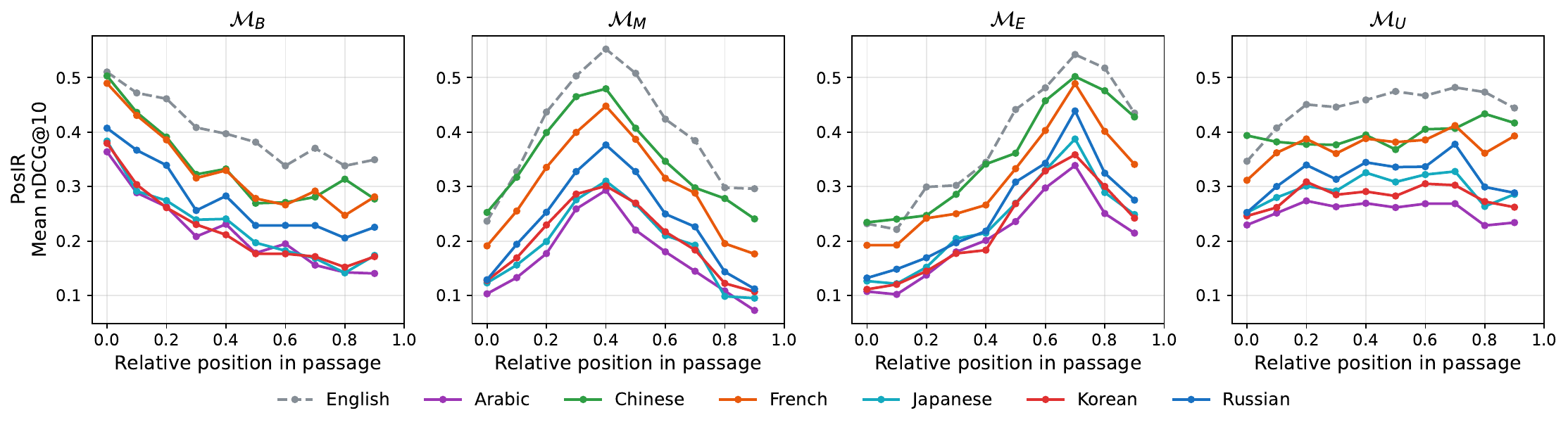}
    % \vspace{-2em}
    \caption{Position-wise nDCG@10 on the four selected PosIR domains (Subject Education, News Media, Law Judiciary, and Finance Economics) for Qwen3-0.6B retrievers fine-tuned only on English position-controlled data and evaluated zero-shot in seven languages. Columns correspond to training configurations $\mathcal{M}_B$, $\mathcal{M}_M$, $\mathcal{M}_E$, and $\mathcal{M}_U$; lines denote evaluation languages. Each skewed configuration retains its trained-position preference in every language, while $\mathcal{M}_U$ remains comparatively flat.}
    \label{fig:posir-multilingual}
\end{figure*}

Figure~\ref{fig:posir-multilingual} shows that the directional pattern
transfers zero-shot: in every evaluation language, $\mathcal{M}_B$ favors
earlier evidence, $\mathcal{M}_M$ peaks near the middle, $\mathcal{M}_E$
improves toward later positions, and $\mathcal{M}_U$ remains comparatively
flat. Absolute nDCG@10 varies across languages, but the position
preference consistently follows the English training distribution.
Table~\ref{tab:posir-multilingual} quantifies this: $\mathcal{M}_U$
attains the highest mean nDCG@10 in all seven languages (+10.0--27.3\%
over the best skewed configuration) together with the lowest PSI
(0.132--0.261 versus 0.332--0.637 for the skewed configurations). Because the evaluation languages
share little surface or lexical overlap with the English training data,
the transferred positional preference is difficult to attribute to
English-specific surface or query-style artifacts; the models instead
appear to encode the position of evidence itself, consistent with the
shortcut account in Section~\ref{sec:mechanism}. Content- and
discourse-level confounds, which multilingual pretraining does not rule
out, are addressed separately by the segment-permutation intervention
(Appendix~\ref{app:permutation}).

\subsection{Human-Written Korean Questions (KorQuAD)}
\label{app:korquad}
To corroborate this transfer with human-written queries, we further
evaluate the same retrievers---together with the unfine-tuned
Qwen3-0.6B checkpoint---on KorQuAD~v1 \citep{korquad}, grouping
its 66,639 human-written Korean questions into six bins by the
character-start offset of the gold answer, following the SQuAD-PosQ
protocol. Table~\ref{tab:korquad} shows the same directional transfer:
the performance peak shifts from early to late bins as the English
training emphasis moves from begin to middle to end, while
$\mathcal{M}_U$ is substantially flatter (PSI 0.167 versus
0.438--0.530). The unfine-tuned base shows near-floor performance across bins, indicating that the
base checkpoint alone has no usable retrieval ability on this task;
the measured pattern therefore emerges only through the English
fine-tuning. Moreover, since pretraining is shared across the four
configurations, the directional differences among them are
attributable to the training-position distribution rather than to
pretraining. Because these questions are human-written and
share no surface or lexical overlap with the English training data,
the transferred prior can be attributed neither to the LLM
query-generation style nor to surface-level English-specific
artifacts.

We note the scope of these results: they establish zero-shot transfer
of an English-induced prior, not evidence about multilingual training.
Whether fine-tuning on position-controlled data in other languages
induces or mitigates the same bias remains future work.

\begin{table}[t]
\centering
\small
\resizebox{\columnwidth}{!}{%
\setlength{\tabcolsep}{3.5pt}
\begin{tabular}{lcccccccc}
\toprule
& \multicolumn{4}{c}{\textbf{NDCG@10}} & \multicolumn{4}{c}{\textbf{PSI} $\downarrow$} \\
\cmidrule(lr){2-5} \cmidrule(lr){6-9}
Language & $\mathcal{M}_B$ & $\mathcal{M}_M$ & $\mathcal{M}_E$ & $\mathcal{M}_U$
         & $\mathcal{M}_B$ & $\mathcal{M}_M$ & $\mathcal{M}_E$ & $\mathcal{M}_U$ \\
\midrule
English  & 0.409 & 0.401 & 0.386 & \textbf{0.450} & 0.341 & 0.562 & 0.590 & \textbf{0.261} \\
Arabic   & 0.222 & 0.172 & 0.205 & \textbf{0.257} & 0.456 & 0.525 & 0.624 & \textbf{0.163} \\
Chinese  & 0.349 & 0.350 & 0.349 & \textbf{0.394} & 0.351 & 0.332 & 0.512 & \textbf{0.175} \\
French   & 0.337 & 0.304 & 0.308 & \textbf{0.374} & 0.351 & 0.412 & 0.535 & \textbf{0.132} \\
Japanese & 0.233 & 0.196 & 0.232 & \textbf{0.297} & 0.447 & 0.499 & 0.612 & \textbf{0.197} \\
Korean   & 0.227 & 0.205 & 0.221 & \textbf{0.283} & 0.450 & 0.478 & 0.637 & \textbf{0.161} \\
Russian  & 0.282 & 0.239 & 0.253 & \textbf{0.321} & 0.401 & 0.481 & 0.604 & \textbf{0.176} \\
\bottomrule
\end{tabular}
}
\caption{Mean nDCG@10 and Position Sensitivity Index (PSI) by evaluation
language for Qwen3-0.6B retrievers fine-tuned only on English
position-controlled data, evaluated zero-shot on the four selected PosIR
domains (Subject Education, News Media, Law Judiciary, and Finance Economics).
Higher is better for nDCG@10; lower is better for PSI. Best values for
each language and metric are in \textbf{bold}.}
\label{tab:posir-multilingual}
\end{table}
\begin{table}[t]
\centering
\small
\resizebox{\columnwidth}{!}{%
\setlength{\tabcolsep}{3.5pt}
\begin{tabular}{lccccccc}
\toprule
\textbf{Config} & \textbf{0+} & \textbf{100+} & \textbf{200+} & \textbf{300+} & \textbf{400+} & \textbf{500+} & \textbf{PSI} $\downarrow$ \\
\midrule
Base            & 0.033 & 0.034 & 0.039 & 0.044 & 0.040 & 0.038 & --- \\
$\mathcal{M}_B$ & \textbf{0.639} & 0.481 & 0.389 & 0.366 & 0.337 & 0.300 & 0.530 \\
$\mathcal{M}_M$ & 0.420 & \textbf{0.607} & 0.576 & 0.434 & 0.390 & 0.341 & 0.438 \\
$\mathcal{M}_E$ & 0.351 & 0.421 & 0.589 & \textbf{0.667} & \textbf{0.654} & \textbf{0.604} & 0.474 \\
$\mathcal{M}_U$ & 0.621 & 0.630 & \textbf{0.634} & 0.602 & 0.584 & 0.528 & \textbf{0.167} \\
\bottomrule
\end{tabular}
}
\caption{Zero-shot transfer to human-written Korean questions.
nDCG@10 on KorQuAD~v1 for Qwen3-0.6B retrievers
fine-tuned only on English position-controlled data; Base denotes the
unfine-tuned checkpoint. Questions are grouped into six bins by the
character-start offset of the gold answer; columns are labeled by the
lower edge of each 100-character bin (0+ denotes offsets in $[0,100)$,
\ldots, and 500+ denotes offsets beyond 500). The highest bin for each configuration and the
lowest PSI are in \textbf{bold}.}
\label{tab:korquad}
\end{table}

% =====================================================================
% 부록 신규 소절: Segment-Permutation (서술 + 표) — 렌더 확인 완료
% 배치 제안: Appendix E.1(mirror-reversal) 옆에 E.2로 — "위치 조작
%   개입 실험"끼리 묶임. 섹션 번호·라벨은 프로젝트 구조에 맞게 조정.
% 요구 패키지: booktabs, multirow
%
% ★ TODO (리버탈 원문과 대조 필요 — 원문 문구로 교체할 자리):
%  (1) 분산 분해의 정확한 정의 — 아래 서술은 "3×3 셀 평균 분산을
%      position / content / 잔여 interaction으로 분해"라는 일반적
%      표현임. 리버탈 응답의 실제 계산 방식(two-way 분해인지, 주변
%      평균 분산 비율인지)과 문구를 맞출 것.
%  (2) "reversed-distractor control" 한 문장 — 리버탈 원문 표현으로.
%  (3) E.1 mirror-reversal 연결 문장 — E.1의 실제 서술과 정합하게.
%  (4) 벤치마크 재현(43.8–93.1% vs 1.3–12.3%)은 현재 범위만 서술.
%      config별 수치가 있으면 작은 표로 승격 가능 — 값 주시면 추가.
%
% 본문 §6 요약 문단 (그대로 붙여넣기용, 3–4문장):
%   To disentangle physical position from content and discourse role,
%   we permute the identical evidence span across beginning/middle/end
%   placements while holding the query and all content fixed
%   (Appendix~\ref{app:permutation}). Physical position accounts for
%   82.7--97.9\% of the variance in cell-mean nDCG@10 for skewed
%   models but at most 2.1\% for $\mathcal{M}_U$, whose behavior is
%   instead driven by content (92.6--99.7\%); the same pattern holds
%   when permuting \textsc{SQuAD-PosQ} and \textsc{FineWeb-PosQ}
%   documents (43.8--93.1\% vs.\ 1.3--12.3\%). Content role is not
%   fully eliminated---it retains a 28.2\% share for Qwen3
%   $\mathcal{M}_B$---but physical position is the dominant driver of
%   the learned preference.
% =====================================================================

\section{Segment Permutation: Physical Position vs.\ Content}
\label{app:permutation}

The position-aware benchmarks vary where evidence occurs across documents, so evidence position naturally co-varies with the content and discourse role of the surrounding segments. To separate the effect of physical position from these content-level factors, we run a controlled intervention in which the identical evidence span is moved across physical positions while everything else is held fixed.

\paragraph{Protocol.}
We use held-out query--evidence pairs from the position-verified pool (15{,}000 pairs; a 3{,}000-pair subsample for Qwen3-0.6B owing to decoder encoding cost). For each pair we construct three variants of the document that place the same evidence segment at the beginning, middle, or end; the surrounding context consists of the same two remaining segments of that document, so content type, discourse function, and segment difficulty are identical across placements, and only the physical position of the evidence changes. A reversed-distractor control imposes an identical permutation burden on every condition, and queries are never modified. % TODO(2): 리버탈 원문 표현으로 교체
This yields a $3{\times}3$ grid of (evidence content role $\times$ physical position) conditions; we compute mean nDCG@10 per cell and decompose the variance of the cell means into the share attributable to physical position, the share attributable to content role, and a residual interaction term. % TODO(1): 분해 방식 정의를 리버탈 문구와 일치시킬 것

\paragraph{Results.}
Table~\ref{tab:permutation} reports the decomposition. For every skewed configuration, physical position dominates: it accounts for 82.7--92.0\% of cell-mean variance for ModernBERT and 66.8--97.9\% for Qwen3, and moving the same evidence into versus out of the trained position shifts nDCG@10 by up to 37.7 points. $\mathcal{M}_U$ is the mirror image: position explains at most 2.1\% of the variance, content role explains 92.6--99.7\%, and the max--min gap collapses to below one point.

\paragraph{Replication on benchmark documents.}
Applying the same permutation to SQuAD-PosQ and FineWeb-PosQ documents reproduces the pattern (Table~\ref{tab:permutation-benchmark}): the position share ranges over 50.2--93.1\% for skewed configurations versus 1.3--12.2\% for $\mathcal{M}_U$. The natural-order check in parentheses preserves the separation, with lower position shares for Qwen3-0.6B on FineWeb-PosQ.

\paragraph{What this shows---and what it does not.}
The intervention isolates the physical placement of a fixed evidence span; it does not eliminate every content-related factor. In particular, content role retains a 28.2\% share for Qwen3 $\mathcal{M}_B$, indicating partial entanglement between position and content signals in that configuration. We therefore interpret these results as showing that physical position is the \emph{dominant} driver of the learned preference, not that content effects are absent. The direction of the effect is consistent with the mirror-reversal analysis in Appendix~\ref{app:posir-mirror}. % TODO(3): E.1 서술과 정합 확인

% \begin{table*}[t]
% \centering
% \small
% \setlength{\tabcolsep}{5pt}
% \renewcommand{\arraystretch}{1.15}
% \begin{tabular}{llccc}
% \toprule
% \textbf{Model} & \textbf{Config} & \textbf{Position share} & \textbf{Content share} & \textbf{Max--min gap (pp)} \\
% \midrule
% \multirow{4}{*}{\shortstack[l]{ModernBERT-base\\{\scriptsize(15{,}000 pairs)}}}
% & $\mathcal{M}_B$ & \textbf{82.7\%} & 16.7\% & 37.7 \\
% & $\mathcal{M}_M$ & \textbf{92.0\%} & 7.4\%  & 24.8 \\
% & $\mathcal{M}_E$ & \textbf{92.0\%} & 5.2\%  & 32.2 \\
% & $\mathcal{M}_U$ & 0.1\%  & \textbf{99.7\%} & 0.7 \\
% \addlinespace[3pt]
% \multirow{4}{*}{\shortstack[l]{Qwen3-0.6B\\{\scriptsize(3{,}000 pairs)}}}
% & $\mathcal{M}_B$ & \textbf{66.8\%} & 28.2\% & 20.8 \\
% & $\mathcal{M}_M$ & \textbf{97.3\%} & 0.5\%  & 17.3 \\
% & $\mathcal{M}_E$ & \textbf{97.9\%} & 0.4\%  & 25.8 \\
% & $\mathcal{M}_U$ & 2.1\%  & \textbf{92.6\%} & 0.9 \\
% \bottomrule
% \end{tabular}
% \caption{Variance decomposition of the segment-permutation intervention. The identical evidence span is placed at the beginning, middle, or end of the document while the query and all content are held fixed. Position and content shares give the fraction of variance in cell-mean nDCG@10 attributable to physical position and to evidence content role, respectively; the remainder is their interaction. Max--min gap is the spread of cell-mean nDCG@10 across placements, in points. The dominant share in each row is in \textbf{bold}.}
% \label{tab:permutation}
% \end{table*}

\begin{table*}[t]
\centering
\small
\setlength{\tabcolsep}{5pt}
\renewcommand{\arraystretch}{1.15}
\begin{tabular}{llccc}
\toprule
\textbf{Model} & \textbf{Config} & \textbf{Position share} & \textbf{Content share} & \textbf{Max--min gap (pp)} \\
\midrule
\multirow{4}{*}{\shortstack[l]{ModernBERT-base\\{\scriptsize(15{,}000 pairs)}}}
& $\mathcal{M}_B$ & \textbf{82.7\%} & 16.7\% & 37.7 \\
& $\mathcal{M}_M$ & \textbf{92.0\%} & 7.4\%  & 24.8 \\
& $\mathcal{M}_E$ & \textbf{92.0\%} & 5.2\%  & 32.2 \\
& $\mathcal{M}_U$ & 0.1\%  & \textbf{99.7\%} & 0.7 \\
\addlinespace[3pt]
\multirow{4}{*}{\shortstack[l]{Qwen3-0.6B\\{\scriptsize(3{,}000 pairs)}}}
& $\mathcal{M}_B$ & \textbf{66.8\%} & 28.2\% & 20.8 \\
& $\mathcal{M}_M$ & \textbf{97.3\%} & 0.5\%  & 17.3 \\
& $\mathcal{M}_E$ & \textbf{97.9\%} & 0.4\%  & 25.8 \\
& $\mathcal{M}_U$ & 2.1\%  & \textbf{92.6\%} & 0.9 \\
\bottomrule
\end{tabular}
\vspace{-0.5em}
\caption{Variance decomposition of the segment-permutation intervention. The identical evidence span is placed at the beginning, middle, or end of the document while the query and all content are held fixed. Position and content shares give the fraction of variance in cell-mean nDCG@10 attributable to physical position and to evidence content role, respectively; the remainder is their interaction. Max--min gap is the spread of cell-mean nDCG@10 across placements, in points. The dominant share in each row is in \textbf{bold}.}
\label{tab:permutation}
\end{table*}
 
\begin{table*}[t]
\centering
\small
\setlength{\tabcolsep}{7pt}
\renewcommand{\arraystretch}{1.15}
\begin{tabular}{llcc}
\toprule
\textbf{Model} & \textbf{Config} & \textbf{\textsc{SQuAD-PosQ}} & \textbf{\textsc{FineWeb-PosQ}} \\
\midrule
\multirow{4}{*}{ModernBERT-base}
& $\mathcal{M}_B$ & 93.1 (80.3) & 91.1 (87.8) \\
& $\mathcal{M}_M$ & 67.8 (71.1) & 90.3 (87.1) \\
& $\mathcal{M}_E$ & 84.8 (89.8) & 89.9 (94.7) \\
& $\mathcal{M}_U$ & 10.0 (14.8) & 10.9 \phantom{0}(9.7) \\
\addlinespace[3pt]
\multirow{4}{*}{Qwen3-0.6B}
& $\mathcal{M}_B$ & 91.0 (79.3) & 77.1 (40.7) \\
& $\mathcal{M}_M$ & 76.2 (79.3) & 69.0 (22.1) \\
& $\mathcal{M}_E$ & 85.3 (89.8) & 50.2 (55.5) \\
& $\mathcal{M}_U$ & \phantom{0}1.3 \phantom{0}(0.5) & 12.2 \phantom{0}(8.8) \\
\bottomrule
\end{tabular}
\vspace{-0.5em}
\caption{Position share (\%) of the variance in cell-mean nDCG@10 when the segment permutation is applied to benchmark documents. Primary values use the constant-burden (reversed-distractor) ordering; parentheses report the natural-order robustness check. Sample sizes: $n{=}17{,}757$ for \textsc{SQuAD-PosQ} (context-deduplicated; the answer segment is located exactly via answer starts) and $n{=}9{,}563$ for \textsc{FineWeb-PosQ} (content-deduplicated; answer segments follow span-class labels).}
\label{tab:permutation-benchmark}
\vspace{-1.5em}
\end{table*}

\section{Evidence-Moving Analysis Full Results}
\label{app:passage-moving}

\begin{table*}[t]
\centering
\small
\resizebox{\linewidth}{!}{%
\begin{tabular}{lc cccccccccc c}
\toprule
\textbf{Base Model} & \textbf{Config} & \textbf{p1} & \textbf{p2} & \textbf{p3} & \textbf{p4} & \textbf{p5} & \textbf{p6} & \textbf{p7} & \textbf{p8} & \textbf{p9} & \textbf{p10} & \textbf{Range} \\
\midrule
\multirow{4}{*}{BERT-base}
 & $\mathcal{M}_B$   & \textbf{.6832} & .6720 & .6777 & .6805 & .6733 & \underline{.6713} & .6730 & .6750 & .6778 & .6803 & 11.9 \\
 & $\mathcal{M}_M$     & .6599 & .6596 & \underline{.6577} & .6686 & \textbf{.6869} & .6732 & .6722 & .6666 & .6613 & .6722 & 29.3 \\
 & $\mathcal{M}_E$     & \underline{.6326} & .6393 & .6371 & .6332 & .6366 & \textbf{.6605} & .6582 & .6547 & .6399 & .6337 & 27.8 \\
 & $\mathcal{M}_U$ & \underline{.6702} & .6717 & .6715 & .6716 & .6739 & .6775 & \textbf{.6796} & .6780 & .6755 & .6746 &  9.4 \\
\midrule
\multirow{4}{*}{ModernBERT-base}
 & $\mathcal{M}_B$   & \textbf{.6582} & .6454 & .6420 & .6403 & .6387 & .6380 & .6375 & .6369 & \underline{.6366} & .6370 & 21.5 \\
 & $\mathcal{M}_M$     & .6045 & .6064 & .6077 & \textbf{.6078} & .6074 & .6062 & .6048 & .6027 & .6008 & \underline{.5985} &  9.4 \\
 & $\mathcal{M}_E$     & \underline{.5622} & .5626 & .5637 & .5653 & .5674 & .5706 & .5759 & .5802 & \textbf{.5829} & .5803 & 20.6 \\
 & $\mathcal{M}_U$ & .6144 & \underline{.6143} & .6147 & .6148 & .6149 & .6148 & .6151 & .6147 & .6150 & \textbf{.6162} &  1.9 \\
\midrule
\multirow{4}{*}{ModernBERT-large}
 & $\mathcal{M}_B$   & \textbf{.6452} & .6364 & .6345 & .6331 & .6327 & .6322 & .6321 & .6318 & \underline{.6315} & .6323 & 13.7 \\
 & $\mathcal{M}_M$     & .6054 & .6062 & \textbf{.6067} & .6064 & .6059 & .6051 & .6046 & .6033 & .6024 & \underline{.6010} &  5.7 \\
 & $\mathcal{M}_E$     & \underline{.5732} & .5747 & .5757 & .5766 & .5777 & .5791 & .5814 & .5834 & .5856 & \textbf{.5881} & 14.9 \\
 & $\mathcal{M}_U$ & \underline{.6160} & .6168 & .6169 & .6165 & .6164 & .6166 & .6169 & .6169 & .6169 & \textbf{.6175} &  1.6 \\
\midrule
\multirow{4}{*}{Longformer-base}
 & $\mathcal{M}_B$   & \textbf{.6261} & .6204 & .6188 & .6174 & .6174 & .6180 & .6192 & .6203 & .6183 & \underline{.6164} &  9.7 \\
 & $\mathcal{M}_M$     & .5838 & .5856 & .5871 & .5877 & \textbf{.5878} & .5870 & .5869 & .5855 & .5849 & \underline{.5807} &  7.1 \\
 & $\mathcal{M}_E$     & \underline{.5622} & .5641 & .5656 & .5660 & .5671 & .5681 & .5691 & .5696 & .5712 & \textbf{.5779} & 15.7 \\
 & $\mathcal{M}_U$ & \underline{.6030} & .6033 & .6043 & .6042 & .6049 & .6050 & .6056 & .6058 & .6057 & \textbf{.6061} &  3.1 \\
\midrule
\multirow{4}{*}{GPT-2-medium}
 & $\mathcal{M}_B$   & \underline{.6637} & .6684 & .6666 & .6655 & .6686 & .6693 & .6658 & .6711 & .6687 & \textbf{.6713} &  7.6 \\
 & $\mathcal{M}_M$     & .6984 & .6989 & .6969 & \underline{.6945} & .6983 & .6994 & .6960 & .7004 & .6979 & \textbf{.7009} &  6.5 \\
 & $\mathcal{M}_E$     & .7044 & .7040 & .7029 & \underline{.7011} & .7023 & .7047 & .7030 & .7049 & .7060 & \textbf{.7109} &  9.8 \\
 & $\mathcal{M}_U$ & .6831 & \textbf{.6876} & .6846 & \underline{.6816} & .6861 & .6868 & .6822 & .6861 & .6851 & .6819 &  6.0 \\
\midrule
\multirow{4}{*}{BLOOM-560M}
 & $\mathcal{M}_B$   & \textbf{.3289} & .3174 & .3164 & .3167 & .3185 & .3146 & .3163 & .3149 & .3143 & \underline{.3058} & 23.1 \\
 & $\mathcal{M}_M$     & .5459 & .5483 & .5487 & \textbf{.5489} & .5482 & .5465 & .5467 & .5458 & .5460 & \underline{.5442} &  4.7 \\
 & $\mathcal{M}_E$     & .5576 & \underline{.5562} & .5608 & .5627 & .5629 & .5601 & .5673 & .5686 & \textbf{.5776} & .5579 & 21.4 \\
 & $\mathcal{M}_U$ & .6026 & .6024 & .6032 & .6037 & .6032 & \underline{.6004} & .6029 & .6034 & .6054 & \textbf{.6058} &  5.4 \\
\midrule
\multirow{4}{*}{TinyLlama-NoPE}
 & $\mathcal{M}_B$   & \textbf{.4401} & .4266 & .4254 & .4238 & .4231 & .4235 & .4228 & .4238 & .4228 & \underline{.4207} & 19.5 \\
 & $\mathcal{M}_M$     & \textbf{.4514} & .4483 & .4479 & .4477 & \underline{.4471} & .4472 & .4476 & .4477 & .4472 & .4378 & 13.6 \\
 & $\mathcal{M}_E$     & .4068 & \underline{.4034} & .4035 & .4042 & .4055 & .4068 & .4081 & .4095 & .4155 & \textbf{.4182} & 14.8 \\
 & $\mathcal{M}_U$ & \textbf{.4363} & .4324 & .4323 & .4322 & .4322 & .4321 & .4326 & .4331 & .4334 & \underline{.4276} &  8.7 \\
\midrule
\multirow{4}{*}{Qwen3-0.6B}
 & $\mathcal{M}_B$   & \textbf{.6085} & .6035 & .6003 & .5986 & .5989 & .5981 & .5986 & .5977 & .5981 & \underline{.5870} & 21.5 \\
 & $\mathcal{M}_M$     & .5401 & .5428 & .5459 & .5489 & \textbf{.5509} & .5484 & .5451 & .5417 & .5397 & \underline{.5238} & 27.1 \\
 & $\mathcal{M}_E$     & \underline{.5161} & .5174 & .5172 & .5177 & .5216 & .5254 & .5309 & .5353 & \textbf{.5368} & .5278 & 20.6 \\
 & $\mathcal{M}_U$ & .5498 & .5519 & .5513 & .5505 & .5518 & .5516 & .5521 & .5516 & \textbf{.5522} & \underline{.5467} &  5.5 \\
\bottomrule
\end{tabular}
}
\vspace{-0.5em}
\caption{Full evidence-moving cosine similarity across all eight models within each document (\textit{p}1--\textit{p}10). \textbf{Bold} marks the highest cosine (peak); \underline{underline} marks the lowest cosine. \textbf{Range} is the cosine difference between peak and lowest ($\times 10^{3}$).}
\label{tab:evidence-moving-full}
\end{table*}

Table~\ref{tab:evidence-moving-full} reports the full position-wise cosine similarity for all eight models under the evidence-moving experiment described in Section~\ref{sec:analyses}, with insertion positions \textit{p}1–\textit{p}10 ordered from earliest to latest.

Five of the eight models---ModernBERT-base, ModernBERT-large, Longformer-base, BLOOM-560M, and Qwen3-0.6B---show clear directional alignment with the fine-tuning distribution: $\mathcal{M}_B$ models peak at \textit{p}1, $\mathcal{M}_E$ models peak at \textit{p}9 or \textit{p}10, and $\mathcal{M}_M$ models peak in the middle range, between \textit{p}3 and \textit{p}5. Uniform training produces the smallest Range for seven of the eight models. The only exception is BLOOM-560M, where the mid-trained Range is 4.7 and the uniform-trained Range is 5.4; both values are much lower than the begin- and end-trained Ranges of 23.1 and 21.4. Overall, the uniform setting compresses the peak-to-lowest cosine differences and weakens position-specific preference.

The remaining three models exhibit model-specific deviations from clean directional alignment. GPT-2-medium shows a persistent late-position preference: cosine similarity peaks at \textit{p}10 under the begin-, mid-, and end-trained configurations, suggesting that fine-tuning does not fully redirect this pre-existing tendency; under uniform training, however, the peak shifts to \textit{p}2 and the Range decreases to 6.0. BERT-base aligns with begin and middle training, peaking at \textit{p}5 under $\mathcal{M}_M$, but its end-trained configuration peaks at \textit{p}6 rather than at \textit{p}9 or \textit{p}10. TinyLlama-NoPE aligns under begin and end training, peaking at \textit{p}1 under $\mathcal{M}_B$ and \textit{p}10 under $\mathcal{M}_E$, but its peak remains at \textit{p}1 under mid-training.

These model-specific deviations do not change the ranking-level conclusions in Section~\ref{sec:results}: across all eight models, position-skewed fine-tuning induces retrieval behavior aligned with the corresponding training-position distribution. Table~\ref{tab:evidence-moving-full} instead shows that the strength and exact embedding-level location of this preference can vary by architecture, with some models retaining residual positional tendencies even after controlled fine-tuning.

\begin{table*}[t]
\centering
\small
\setlength{\tabcolsep}{2.5pt}
\renewcommand{\arraystretch}{1.15}
\begin{tabular}{lcccccccc}
\toprule
\textbf{Config} & \textbf{0--100} & \textbf{100--200} & \textbf{200--300} & \textbf{300--400} & \textbf{400--500} & \textbf{500+} & \textbf{Mean} & \textbf{PSI} $\downarrow$ \\
\midrule
\multicolumn{9}{l}{\textit{ModernBERT-base}} \\
\addlinespace[1pt]
$\mathcal{M}_B$ & \textbf{0.589}{\scriptsize$\pm$0.004} & \textbf{0.565}{\scriptsize$\pm$0.011} & 0.511{\scriptsize$\pm$0.014} & 0.454{\scriptsize$\pm$0.016} & 0.417{\scriptsize$\pm$0.018} & 0.352{\scriptsize$\pm$0.018} & 0.481{\scriptsize$\pm$0.013} & 0.403{\scriptsize$\pm$0.026} \\
$\mathcal{M}_M$ & 0.463{\scriptsize$\pm$0.005} & 0.529{\scriptsize$\pm$0.001} & \textbf{0.584}{\scriptsize$\pm$0.001} & \textbf{0.584}{\scriptsize$\pm$0.002} & 0.539{\scriptsize$\pm$0.002} & 0.415{\scriptsize$\pm$0.005} & 0.519{\scriptsize$\pm$0.002} & 0.289{\scriptsize$\pm$0.009} \\
$\mathcal{M}_E$ & 0.377{\scriptsize$\pm$0.007} & 0.400{\scriptsize$\pm$0.006} & 0.434{\scriptsize$\pm$0.004} & 0.488{\scriptsize$\pm$0.006} & \textbf{0.549}{\scriptsize$\pm$0.005} & \textbf{0.566}{\scriptsize$\pm$0.005} & 0.469{\scriptsize$\pm$0.005} & 0.334{\scriptsize$\pm$0.007} \\
$\mathcal{M}_U$ & 0.492{\scriptsize$\pm$0.005} & 0.518{\scriptsize$\pm$0.005} & 0.535{\scriptsize$\pm$0.007} & 0.538{\scriptsize$\pm$0.004} & 0.536{\scriptsize$\pm$0.001} & 0.503{\scriptsize$\pm$0.002} & \textbf{0.520}{\scriptsize$\pm$0.004} & \textbf{0.085}{\scriptsize$\pm$0.006} \\
\midrule
\multicolumn{9}{l}{\textit{Qwen3-0.6B}} \\
\addlinespace[1pt]
$\mathcal{M}_B$ & \textbf{0.688}{\scriptsize$\pm$0.016} & 0.645{\scriptsize$\pm$0.013} & 0.574{\scriptsize$\pm$0.015} & 0.514{\scriptsize$\pm$0.018} & 0.475{\scriptsize$\pm$0.022} & 0.399{\scriptsize$\pm$0.025} & 0.549{\scriptsize$\pm$0.016} & 0.420{\scriptsize$\pm$0.036} \\
$\mathcal{M}_M$ & 0.549{\scriptsize$\pm$0.007} & \textbf{0.649}{\scriptsize$\pm$0.004} & \textbf{0.717}{\scriptsize$\pm$0.001} & \textbf{0.703}{\scriptsize$\pm$0.005} & 0.636{\scriptsize$\pm$0.004} & 0.514{\scriptsize$\pm$0.001} & 0.628{\scriptsize$\pm$0.002} & 0.282{\scriptsize$\pm$0.001} \\
$\mathcal{M}_E$ & 0.402{\scriptsize$\pm$0.015} & 0.442{\scriptsize$\pm$0.016} & 0.497{\scriptsize$\pm$0.016} & 0.574{\scriptsize$\pm$0.014} & 0.663{\scriptsize$\pm$0.009} & \textbf{0.704}{\scriptsize$\pm$0.005} & 0.547{\scriptsize$\pm$0.012} & 0.429{\scriptsize$\pm$0.020} \\
$\mathcal{M}_U$ & 0.615{\scriptsize$\pm$0.016} & 0.642{\scriptsize$\pm$0.010} & 0.659{\scriptsize$\pm$0.007} & 0.667{\scriptsize$\pm$0.003} & \textbf{0.672}{\scriptsize$\pm$0.003} & 0.630{\scriptsize$\pm$0.001} & \textbf{0.648}{\scriptsize$\pm$0.006} & \textbf{0.088}{\scriptsize$\pm$0.018} \\
\bottomrule
\end{tabular}
\caption{Three-seed replication of the main results (Table~\ref{tab:main-results})
on \textsc{SQuAD-PosQ} for ModernBERT-base and Qwen3-0.6B. Each run independently
re-draws the training sets from the position-verified pool (data seeds 42, 84,
100), with the training seed held fixed. The reported variability
therefore reflects training-data re-sampling together with the induced change in
the training trajectory. The single-run patterns of Table~\ref{tab:main-results}
replicate across seeds: each skewed configuration peaks at its trained position,
and $\mathcal{M}_U$ remains near-flat, combining the best overall retrieval
quality with the greatest positional stability on both backbones.
Within each model, the highest nDCG@10 in each bin, the
highest Mean, and the lowest PSI are in \textbf{bold}.}
\label{tab:seed-replication-squad}
\end{table*}

\begin{table*}[t]
\centering
\small
\setlength{\tabcolsep}{5pt}
\renewcommand{\arraystretch}{1.15}
\begin{tabular}{lccccc}
\toprule
\textbf{Config} & \textbf{Begin} & \textbf{Mid} & \textbf{End} & \textbf{Mean} & \textbf{PSI} $\downarrow$ \\
\midrule
\multicolumn{6}{l}{\textit{ModernBERT-base}} \\
\addlinespace[1pt]
$\mathcal{M}_B$ & \textbf{0.780}{\scriptsize$\pm$0.012} & 0.505{\scriptsize$\pm$0.010} & 0.409{\scriptsize$\pm$0.007} & 0.565{\scriptsize$\pm$0.009} & 0.475{\scriptsize$\pm$0.003} \\
$\mathcal{M}_M$ & 0.559{\scriptsize$\pm$0.008} & \textbf{0.734}{\scriptsize$\pm$0.002} & 0.423{\scriptsize$\pm$0.002} & 0.572{\scriptsize$\pm$0.002} & 0.424{\scriptsize$\pm$0.002} \\
$\mathcal{M}_E$ & 0.456{\scriptsize$\pm$0.018} & 0.562{\scriptsize$\pm$0.009} & \textbf{0.688}{\scriptsize$\pm$0.009} & 0.569{\scriptsize$\pm$0.007} & 0.338{\scriptsize$\pm$0.031} \\
$\mathcal{M}_U$ & 0.649{\scriptsize$\pm$0.006} & 0.677{\scriptsize$\pm$0.003} & 0.603{\scriptsize$\pm$0.001} & \textbf{0.643}{\scriptsize$\pm$0.003} & \textbf{0.110}{\scriptsize$\pm$0.004} \\
\midrule
\multicolumn{6}{l}{\textit{Qwen3-0.6B}} \\
\addlinespace[1pt]
$\mathcal{M}_B$ & \textbf{0.749}{\scriptsize$\pm$0.012} & 0.546{\scriptsize$\pm$0.016} & 0.482{\scriptsize$\pm$0.011} & 0.592{\scriptsize$\pm$0.012} & 0.356{\scriptsize$\pm$0.012} \\
$\mathcal{M}_M$ & 0.548{\scriptsize$\pm$0.008} & \textbf{0.615}{\scriptsize$\pm$0.013} & 0.471{\scriptsize$\pm$0.015} & 0.545{\scriptsize$\pm$0.010} & 0.235{\scriptsize$\pm$0.009} \\
$\mathcal{M}_E$ & 0.464{\scriptsize$\pm$0.015} & 0.484{\scriptsize$\pm$0.014} & \textbf{0.668}{\scriptsize$\pm$0.024} & 0.539{\scriptsize$\pm$0.014} & 0.304{\scriptsize$\pm$0.029} \\
$\mathcal{M}_U$ & 0.639{\scriptsize$\pm$0.003} & 0.568{\scriptsize$\pm$0.001} & 0.605{\scriptsize$\pm$0.007} & \textbf{0.604}{\scriptsize$\pm$0.003} & \textbf{0.111}{\scriptsize$\pm$0.004} \\
\bottomrule
\end{tabular}
\caption{Three-seed replication on \textsc{FineWeb-PosQ}, following the same
protocol as Table~\ref{tab:seed-replication-squad}. The single-run patterns of
Table~\ref{tab:main-results} again replicate: every skewed configuration peaks at
its trained position in each individual seed, while $\mathcal{M}_U$ stays
near-flat and delivers the strongest average retrieval with the least positional
sensitivity on both backbones. Within each model, the highest nDCG@10 in each
position, the highest Mean, and the lowest PSI are in \textbf{bold}.}
\label{tab:seed-replication-fineweb}
\end{table*}

\section{Seed Replication}
\label{app:seed-replication}
 
We replicate the main results for ModernBERT-base and Qwen3-0.6B, covering one encoder and one decoder backbone. Each replication uses a distinct data seed (42, 84, 100) that re-runs the data-preparation stage on the position-verified pool: the internal held-out split is re-sampled, the per-cell sampling budget is recomputed, and the training set of every configuration is independently re-drawn. The training seed and all evaluation benchmarks are identical across runs, so the reported variability isolates training-data re-sampling together with the training trajectory it induces. Tables~\ref{tab:seed-replication-squad} and \ref{tab:seed-replication-fineweb} report the results: the single-run patterns of Table~\ref{tab:main-results} replicate across all seeds.

\section{Downstream RAG Evaluation}
\label{app:rag}

To test whether retrieval-level position bias propagates to downstream QA, we build a standard RAG pipeline on \textsc{SQuAD-PosQ} that holds the generator (Qwen3-8B, greedy decoding), prompt, corpus, and context ordering fixed and swaps only the retriever among the four Qwen3-0.6B configurations ($\mathcal{M}_B$, $\mathcal{M}_M$, $\mathcal{M}_E$, $\mathcal{M}_U$), so that any difference in answer quality is attributable to retrieval. We sample 2{,}000 questions from each of the six answer-start bins (100-character bins from 0 to 500, plus a 500${+}$ bin; 11{,}987 unique questions; seed 42); each retriever selects the top-$1$ or top-$5$ of the 20{,}233 benchmark contexts, and answers are scored with the official SQuAD metrics (Table~\ref{tab:rag-endtask}).

Per-bin QA accuracy mirrors each retriever's training bias, and position-balanced retrieval largely removes this sensitivity: QA-PSI falls from 0.448 (worst skewed) to 0.104 at top-$1$ and from 0.335 to 0.055 at top-$5$, a 77--84\% reduction paralleling the retrieval-level reductions, and $\mathcal{M}_U$ attains the highest minimum and bin-averaged scores throughout. Three observations support a retrieval-driven account: per-bin F1 rises almost linearly with per-bin recall ($r = 0.982$ at top-$1$, $0.974$ at top-$5$); the no-context baseline is low and only weakly position-dependent, ruling out the generator's parametric knowledge as the source of the pattern; and on retrieval misses F1 collapses to 12.0--14.4, below even the 16.9--19.3 the same questions attain with no context,\footnote{A miss means no gold context appears in the retrieved top-$k$ ($n_{\text{miss}} = 3{,}039$--$7{,}087$ per condition); the no-context figures are computed on the same miss subsets and hence differ from the full no-context row in Table~\ref{tab:rag-endtask}.} indicating abstention rather than compensation. All $\mathcal{M}_U$-versus-skewed comparisons are significant (McNemar on EM, $p \le 2.3{\times}10^{-8}$; paired-bootstrap F1 gains of ${+}2.36$ to ${+}8.07$ points, $p < 10^{-4}$). We do not claim that balanced training maximizes mean accuracy on every distribution; the finding is that retrieval-level positional instability propagates downstream, and that position-balanced training confines this failure mode.

\begin{table*}[t]
\centering
\small
\setlength{\tabcolsep}{2.2pt}
\renewcommand{\arraystretch}{1.12}
\begin{tabular}{cl *{7}{c} c | *{7}{c} c}
\toprule
& & \multicolumn{8}{c}{\textbf{\textit{Top-1}}} & \multicolumn{8}{c}{\textbf{\textit{Top-5}}} \\
\cmidrule(lr){3-10}\cmidrule(lr){11-18}
& \textbf{Config} & 0+ & 100+ & 200+ & 300+ & 400+ & 500+ & \textbf{Mean} & \textbf{PSI} $\downarrow$ & 0+ & 100+ & 200+ & 300+ & 400+ & 500+ & \textbf{Mean} & \textbf{PSI} $\downarrow$ \\
\midrule
\multirow{4}{*}{\rotatebox[origin=c]{90}{\textit{Recall}}} & $\mathcal{M}_B$ & \textbf{53.6} & 50.9 & 43.6 & 37.4 & 35.2 & 25.8 & 41.1 & 0.519 & \textbf{75.3} & 71.6 & 66.0 & 59.3 & 57.3 & 46.5 & 62.7 & 0.382 \\
 & $\mathcal{M}_M$ & 38.6 & 51.3 & \textbf{58.3} & \textbf{53.6} & 49.3 & 34.2 & 47.6 & 0.414 & 63.8 & 74.5 & \textbf{80.3} & \textbf{77.0} & 72.4 & 57.9 & 71.0 & 0.279 \\
 & $\mathcal{M}_E$ & 25.2 & 30.8 & 35.4 & 43.5 & 52.6 & \textbf{57.8} & 40.9 & 0.564 & 49.7 & 55.6 & 59.4 & 66.5 & 75.0 & \textbf{79.2} & 64.2 & 0.372 \\
 & $\mathcal{M}_U$ & 45.8 & \textbf{52.1} & 53.0 & 52.2 & \textbf{54.0} & 48.3 & \textbf{50.9} & \textbf{0.153} & 73.0 & \textbf{75.4} & 76.5 & 75.2 & \textbf{75.5} & 72.3 & \textbf{74.7} & \textbf{0.055} \\
\midrule
\multirow{5}{*}{\rotatebox[origin=c]{90}{\textit{EM}}} & No context & 10.5 & \phantom{0}8.9 & \phantom{0}8.4 & \phantom{0}7.8 & \phantom{0}7.2 & \phantom{0}8.6 & \phantom{0}8.6 & --- & 10.5 & \phantom{0}8.9 & \phantom{0}8.4 & \phantom{0}7.8 & \phantom{0}7.2 & \phantom{0}8.6 & \phantom{0}8.6 & --- \\
 & $\mathcal{M}_B$ & \textbf{41.0} & 37.5 & 31.8 & 27.5 & 25.9 & 20.1 & 30.6 & 0.510 & \textbf{51.8} & 46.6 & 43.2 & 38.9 & 35.9 & 31.6 & 41.3 & 0.390 \\
 & $\mathcal{M}_M$ & 31.1 & 38.0 & \textbf{41.0} & \textbf{38.2} & 35.0 & 25.7 & 34.8 & 0.374 & 45.4 & \textbf{48.4} & \textbf{52.3} & \textbf{47.5} & 45.2 & 37.9 & 46.1 & 0.276 \\
 & $\mathcal{M}_E$ & 23.9 & 25.2 & 27.4 & 32.3 & 36.5 & \textbf{39.1} & 30.7 & 0.387 & 37.5 & 37.0 & 39.3 & 42.6 & 46.5 & \textbf{48.7} & 41.9 & 0.241 \\
 & $\mathcal{M}_U$ & 36.6 & \textbf{38.6} & 37.3 & 36.9 & \textbf{37.3} & 34.2 & \textbf{36.8} & \textbf{0.112} & 49.9 & \textbf{48.4} & 50.0 & 47.2 & \textbf{46.9} & 46.0 & \textbf{48.1} & \textbf{0.079} \\
\midrule
\multirow{5}{*}{\rotatebox[origin=c]{90}{\textit{F1}}} & No context & 22.9 & 19.7 & 19.5 & 17.9 & 19.1 & 19.4 & 19.8 & --- & 22.9 & 19.7 & 19.5 & 17.9 & 19.1 & 19.4 & 19.8 & --- \\
 & $\mathcal{M}_B$ & \textbf{52.3} & 49.1 & 42.2 & 37.6 & 36.7 & 28.9 & 41.1 & 0.448 & \textbf{64.9} & 60.7 & 56.0 & 51.3 & 49.6 & 43.2 & 54.3 & 0.335 \\
 & $\mathcal{M}_M$ & 41.7 & 49.4 & \textbf{53.2} & \textbf{49.4} & 46.9 & 35.6 & 46.0 & 0.331 & 58.5 & 62.6 & \textbf{66.2} & \textbf{62.4} & 59.6 & 50.9 & 60.0 & 0.231 \\
 & $\mathcal{M}_E$ & 33.0 & 34.8 & 36.9 & 43.0 & 49.0 & \textbf{51.7} & 41.4 & 0.362 & 49.3 & 49.7 & 51.5 & 55.9 & 61.1 & \textbf{64.4} & 55.3 & 0.234 \\
 & $\mathcal{M}_U$ & 47.5 & \textbf{50.7} & 49.4 & 48.6 & \textbf{50.1} & 45.4 & \textbf{48.6} & \textbf{0.104} & 63.5 & \textbf{63.7} & 63.6 & 61.8 & \textbf{61.6} & 60.1 & \textbf{62.4} & \textbf{0.055} \\
\midrule
\multirow{5}{*}{\rotatebox[origin=c]{90}{\textit{Contains}}} & No context & 17.3 & 14.6 & 15.8 & 13.4 & 12.7 & 14.4 & 14.7 & --- & 17.3 & 14.6 & 15.8 & 13.4 & 12.7 & 14.4 & 14.7 & --- \\
 & $\mathcal{M}_B$ & \textbf{51.9} & 47.9 & 40.8 & 35.9 & 33.8 & 26.3 & 39.4 & 0.494 & \textbf{65.6} & 60.2 & 54.8 & 50.0 & 47.1 & 42.4 & 53.4 & 0.353 \\
 & $\mathcal{M}_M$ & 40.1 & 47.2 & \textbf{52.9} & \textbf{48.4} & 44.0 & 33.6 & 44.4 & 0.366 & 57.6 & 61.8 & \textbf{65.8} & \textbf{61.7} & 57.4 & 49.9 & 59.0 & 0.242 \\
 & $\mathcal{M}_E$ & 30.6 & 32.1 & 34.8 & 40.9 & 46.7 & \textbf{50.4} & 39.3 & 0.394 & 48.1 & 48.5 & 50.1 & 54.6 & 58.8 & \textbf{63.9} & 54.0 & 0.246 \\
 & $\mathcal{M}_U$ & 46.3 & \textbf{49.0} & 48.4 & 47.0 & \textbf{46.9} & 44.4 & \textbf{47.0} & \textbf{0.095} & 63.6 & \textbf{62.7} & 62.8 & 61.2 & \textbf{59.3} & 60.1 & \textbf{61.6} & \textbf{0.068} \\
\bottomrule
\end{tabular}
\caption{RAG performance on \textsc{SQuAD-PosQ} with top-$1$ and top-$5$ retrieval, grouped by metric. Columns are labeled by the lower edge of each 100-character answer-start bin ($0{+}$ denotes positions in $[0,100)$, \ldots, and $500{+}$ denotes positions beyond 500). Recall is per-bin retrieval recall, and EM, F1, and contains-answer use the official SQuAD scoring (all in \%). The no-context baseline generates answers without retrieval, isolating the generator's parametric knowledge; it is therefore identical across $k$ and has no recall row. QA-PSI applies the paper's PSI to the per-bin scores and is reported only for the four retriever configurations. Within each metric and each $k$, the highest per-bin score, the highest Mean, and the lowest QA-PSI among the four retrievers are in \textbf{bold}.}
\label{tab:rag-endtask}
\end{table*}

\begin{figure}[t]
    \centering
    \includegraphics[width=\columnwidth]{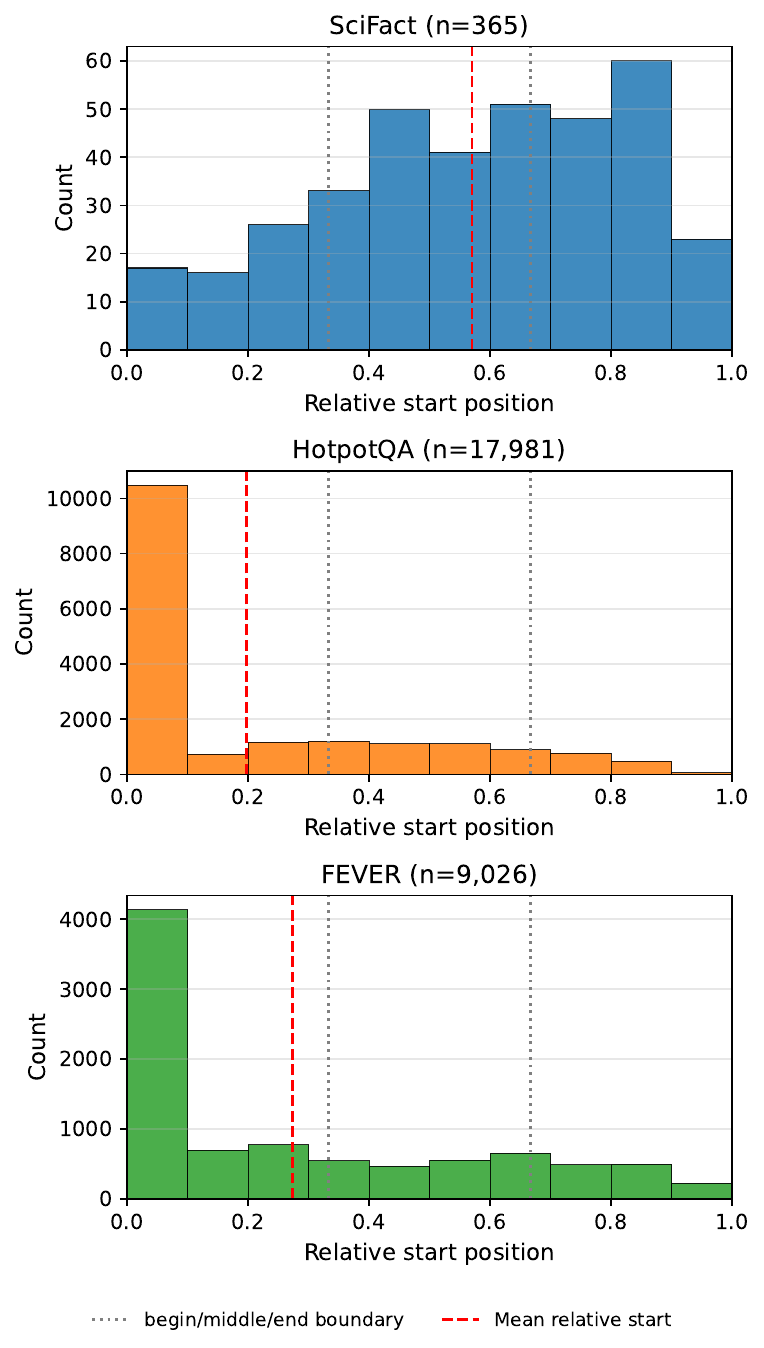}
    \caption{Relative evidence start-position distributions for the BEIR subsets used in our analysis, restricted to examples where the evidence location could be identified. Evidence start positions are normalized by relevant-document length. Dashed red lines indicate mean relative start positions, and dotted gray lines mark the begin/middle/end boundaries.}
    \label{fig:beir-evidence-start}
\end{figure}

\begin{table}[t]
\centering
\small
\setlength{\tabcolsep}{5pt}
\begin{tabular}{lrrrr}
\toprule
BEIR subset & $\mathcal{M}_B$ & $\mathcal{M}_M$ & $\mathcal{M}_E$ & $\mathcal{M}_U$ \\
\midrule
\multicolumn{5}{l}{\textsc{BERT-base}} \\[2pt]
SciFact & \textbf{0.298} & 0.154 & 0.138 & \underline{0.273} \\
HotpotQA & \textbf{0.271} & 0.110 & 0.077 & \underline{0.190} \\
FEVER & \textbf{0.432} & 0.088 & 0.047 & \underline{0.225} \\
\midrule
Average & \textbf{0.334} & 0.117 & 0.087 & \underline{0.229} \\
\midrule
\multicolumn{5}{l}{\textsc{ModernBERT-base}} \\[2pt]
SciFact & \underline{0.422} & 0.379 & 0.383 & \textbf{0.434} \\
HotpotQA & \textbf{0.402} & 0.251 & 0.209 & \underline{0.335} \\
FEVER & \textbf{0.604} & 0.199 & 0.200 & \underline{0.443} \\
\midrule
Average & \textbf{0.476} & 0.276 & 0.264 & \underline{0.404} \\
\midrule
\multicolumn{5}{l}{\textsc{ModernBERT-large}} \\[2pt]
SciFact & 0.472 & \textbf{0.499} & 0.440 & \underline{0.497} \\
HotpotQA & \textbf{0.432} & 0.254 & 0.224 & \underline{0.354} \\
FEVER & \textbf{0.660} & 0.259 & 0.229 & \underline{0.497} \\
\midrule
Average & \textbf{0.521} & 0.337 & 0.298 & \underline{0.449} \\
\midrule
\multicolumn{5}{l}{\textsc{Longformer-base}} \\[2pt]
SciFact & \textbf{0.371} & \underline{0.370} & 0.319 & \underline{0.370} \\
HotpotQA & \textbf{0.408} & 0.255 & 0.253 & \underline{0.358} \\
FEVER & \textbf{0.559} & 0.235 & 0.297 & \underline{0.470} \\
\midrule
Average & \textbf{0.446} & 0.287 & 0.290 & \underline{0.399} \\
\midrule
\multicolumn{5}{l}{\textsc{GPT-2-medium}} \\[2pt]
SciFact & 0.084 & \textbf{0.148} & \underline{0.122} & 0.119 \\
HotpotQA & \textbf{0.096} & 0.050 & 0.031 & \underline{0.053} \\
FEVER & \textbf{0.213} & 0.049 & 0.030 & \underline{0.086} \\
\midrule
Average & \textbf{0.131} & 0.082 & 0.061 & \underline{0.086} \\
\midrule
\multicolumn{5}{l}{\textsc{Qwen3-0.6B}} \\[2pt]
SciFact & 0.550 & 0.604 & \textbf{0.628} & \underline{0.624} \\
HotpotQA & \textbf{0.485} & 0.299 & 0.299 & \underline{0.481} \\
FEVER & \textbf{0.657} & 0.286 & 0.296 & \underline{0.577} \\
\midrule
Average & \textbf{0.564} & 0.396 & 0.408 & \underline{0.561} \\
\midrule
\multicolumn{5}{l}{\textsc{BLOOM-560M}} \\[2pt]
SciFact & 0.365 & \underline{0.458} & 0.384 & \textbf{0.470} \\
HotpotQA & \textbf{0.331} & 0.133 & 0.098 & \underline{0.251} \\
FEVER & \textbf{0.508} & 0.082 & 0.082 & \underline{0.346} \\
\midrule
Average & \textbf{0.401} & 0.224 & 0.188 & \underline{0.356} \\
\midrule
\multicolumn{5}{l}{\textsc{TinyLlama-1.1B}} \\[2pt]
SciFact & 0.243 & \underline{0.329} & 0.308 & \textbf{0.354} \\
HotpotQA & \textbf{0.280} & 0.183 & 0.129 & \underline{0.254} \\
FEVER & \textbf{0.291} & 0.116 & 0.069 & \underline{0.213} \\
\midrule
Average & \underline{0.271} & 0.209 & 0.169 & \textbf{0.274} \\
\bottomrule
\end{tabular}
\caption{Full BEIR nDCG@10 by base model and training configuration. Averages are computed over the three BEIR subsets. Higher is better. Best values within each model--subset row are in bold; second-best values are underlined.}
\label{tab:beir-full-results}
\end{table}

\end{document}